\documentclass[twocolumn]{aastex63}

\shorttitle{Orbital Dynamics in the AU Mic System}
\shortauthors{Stephen R. Kane et al.}

\begin{document}

\title{Orbital Dynamics and the Evolution of Planetary Habitability in
  the AU Mic System}

\author[0000-0002-7084-0529]{Stephen R. Kane}
\affiliation{Department of Earth and Planetary Sciences, University of
  California, Riverside, CA 92521, USA}
\email{skane@ucr.edu}

\author[0000-0002-6943-3192]{Bradford J. Foley}
\affiliation{Department of Geosciences, Pennsylvania State University,
  University Park, PA 16802, USA}

\author[0000-0002-0139-4756]{Michelle L. Hill}
\affiliation{Department of Earth and Planetary Sciences, University of
  California, Riverside, CA 92521, USA}

\author[0000-0001-8991-3110]{Cayman T. Unterborn}
\affiliation{Southwest Research Institute, San Antonio, TX 78238, USA}

\author[0000-0001-7139-2724]{Thomas Barclay}
\affiliation{University of Maryland, Baltimore, MD 21250, USA}
\affiliation{NASA Goddard Space Flight Center, Greenbelt, MD 20771,
  USA}

\author[0000-0001-6279-0595]{Bryson Cale}
\affiliation{Department of Physics \& Astronomy, George Mason
  University, Fairfax, VA 22030, USA}

\author[0000-0002-0388-8004]{Emily A. Gilbert}
\affiliation{Department of Astronomy and Astrophysics, University of
  Chicago, Chicago, IL 60637, USA}
\affiliation{University of Maryland, Baltimore County, Baltimore, MD
  21250, USA}
\affiliation{The Adler Planetarium, Chicago, IL 60605, USA}
\affiliation{NASA Goddard Space Flight Center, Greenbelt, MD 20771,
  USA}
\affiliation{GSFC Sellers Exoplanet Environments Collaboration}

\author[0000-0002-8864-1667]{Peter Plavchan}
\affiliation{Department of Physics \& Astronomy, George Mason
  University, Fairfax, VA 22030, USA}

\author[0000-0002-7424-9891]{Justin M. Wittrock}
\affiliation{Department of Physics \& Astronomy, George Mason
  University, Fairfax, VA 22030, USA}


\begin{abstract}

The diversity of planetary systems that have been discovered are
revealing the plethora of possible architectures, providing insights
into planet formation and evolution. They also increase our
understanding of system parameters that may affect planetary
habitability, and how such conditions are influenced by initial
conditions. The AU~Mic system is unique among known planetary systems
in that it is a nearby, young, multi-planet transiting system. Such a
young and well characterized system provides an opportunity to study
orbital dynamical and habitability studies for planets in the very
early stages of their evolution. Here, we calculate the evolution of
the Habitable Zone of the system through time, including the pre-main
sequence phase that the system currently resides in. We discuss the
planetary atmospheric processes occurring for an Earth-mass planet
during this transitionary period, and provide calculations of the
climate state convergence age for both volatile rich and poor initial
conditions. We present results of an orbital dynamical analysis of the
AU~Mic system that demonstrate the rapid eccentricity evolution of the
known planets, and show that terrestrial planets within the Habitable
Zone of the system can retain long-term stability. Finally, we discuss
follow-up observation prospects, detectability of possible Habitable
Zone planets, and how the AU~Mic system may be used as a template for
studies of planetary habitability evolution.

\end{abstract}

\keywords{astrobiology -- planetary systems -- planets and satellites:
  dynamical evolution and stability -- stars: individual
  (AU~Microscopii)}


\section{Introduction}
\label{intro}

The evolution of surface conditions on terrestrial planets has
numerous driving factors, including the orbital dynamics within the
system and the impact of the host star and planetary interior on the
atmosphere \citep{kasting2003,lammer2009a}. These particular factors
can undergo dramatic changes during the early history of the planetary
system as the star moves through a period of early evolution and
activity \citep{baraffe2002}, the planets interact with each other and
remaining formation material \citep{raymond2012}, and the planetary
interiors contribute to the atmospheric inventory via degassing
\citep{gaillard2014}. Discoveries of young planetary systems thus
provide insight into the conditions present during this rapidly
changing period of planetary system evolution, especially those
systems that are relatively nearby. Examples of nearby young ($<
50$~Myrs) systems with transiting planets include V1298~Tau
\citep{david2019b,david2019c}, located at a distance of $\sim$108~pcs,
and DS~Tuc \citep{newton2019}, located at a distance of
$\sim$44~pcs. These systems provide excellent opportunities for
follow-up observations, such as direct imaging observations that could
extract further properties of the young planets
\citep{davies1980,kane2018c}.

One of the closest young planetary systems currently known is the case
of AU~Microscopii (hereafter AU~Mic). The star is located at a
distance of only 9.79~pcs, has an age estimate of $\sim$22~Myrs
\citep{mamajek2014}, is a famous flare star \citep{plavchan2020}, and
is well known to harbor a substantial debris disk
\cite{strubbe2006,macgregor2013}. A planet orbiting the star was
discovered by \citet{plavchan2020} using photometry from the
Transiting Exoplanet Survey Satellite (TESS; \citet{ricker2015}) and
radial velocities (RV) acquired using the iSHELL \citep{cale2019},
HARPS \citep{pepe2000}, and HIRES \citep{vogt1994} instruments. The
planet has a radius of 0.4~$R_J$, a mass of 0.18~$M_J$, and an orbital
period of 8.46~days. An additional planet was detected in an orbital
period of 18.86~days by \citet{cale2021} via further transit and RV
observations. The relative youth of this system serves as a laboratory
for exploring the early dynamical stability of orbits, and can inform
the viability of planetary orbits in an evolving Habitable Zone (HZ).

In this paper, we present the results of a dynamical analysis of the
AU~Mic system, and a study of the HZ evolution and potential planets
within that region. This system is not only young, but nearby, and is
therefore an excellent target for numerous observational follow-up
studies in addition to those that have been carried out. In
particular, the prospect of eventual terrestrial planet detections at
stable orbital locations would serve as a benchmark for providing a
data-driven evaluation of concepts regarding early terrestrial planet
formation and the evolution of their atmospheres. Section~\ref{system}
describes the system properties and the extent of the HZ at the
present epoch. In Section~\ref{evolution} we provide calculations of
the HZ evolution through time based on stellar isochrones for the host
star, and discuss planetary degassing models during the short-term and
long-term history of the system using various initial condition
assumptions. Section~\ref{dynamics} provides the results of a
dynamical analysis of the known planets, and a detailed study of
possible viable orbits for terrestrial planets at farther distances,
including locations within the HZ. In Section~\ref{conclusions} we
discuss the implications of the system evolution and opportunities for
follow-up observations and studies, and provide concluding remarks.


\section{System Properties and Habitable Zone}
\label{system}

As one of the youngest known exoplanet hosts, the properties of the
host star have been the subject of numerous studies
\citep[e.g.,][]{mamajek2014,kochukhov2020b}. We adopt the stellar
properties provided by \citet{plavchan2020}, including the stellar
effective temperature of $T_\mathrm{eff} = 3700$~K, luminosity of
$L_\star = 0.09$~$L_\odot$, and mass of $M_\star =
0.5$~$M_\odot$. Note that the luminosity is relatively high due to the
pre-main sequence evolutionary stage of the star, discussed in more
detail in Section~\ref{evolution}. The system is known to contain two
planets, b and c, described in detail by \citet{cale2021}. Planet b
has an orbital period of $P_b = 8.46$~days, semi-major axis of $a_b =
0.0645$~AU, eccentricity of $e_b = 0.186$, and planet mass of $M_{p,b}
= 20.12$~$M_\oplus$. Planet c has an orbital period of $P_c =
18.86$~days, semi-major axis of $a_c = 0.1101$~AU, planet mass of
$M_{p,c} = 9.60$~$M_\oplus$, and a circular orbit is assumed.

\begin{figure}
  \includegraphics[angle=270,width=8.5cm]{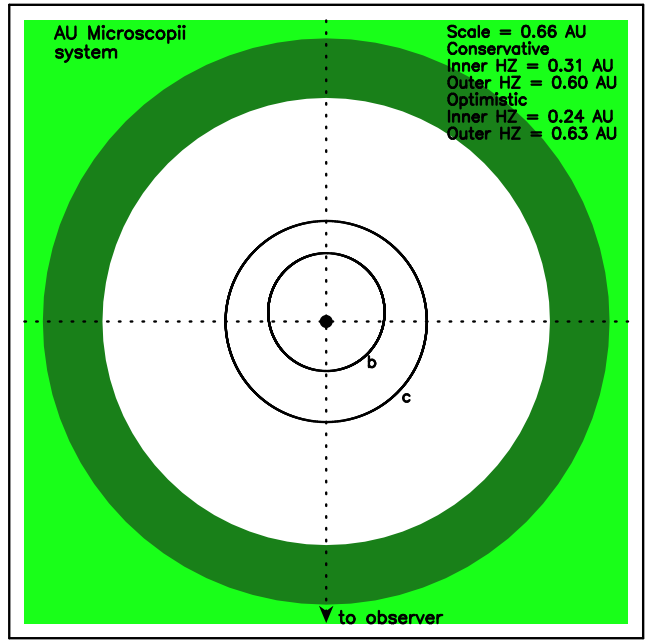}
  \caption{A top-down view of the AU~Mic system with respect to the
    HZ. The star is located at the center of the crosshairs and the
    orbits of the known planets are shown. The CHZ and OHZ are
    indicated by the light green and dark green regions,
    respectively.}
  \label{fig:hz}
\end{figure}

\begin{figure*}
  \begin{center}
    \includegraphics[angle=270,width=16.0cm]{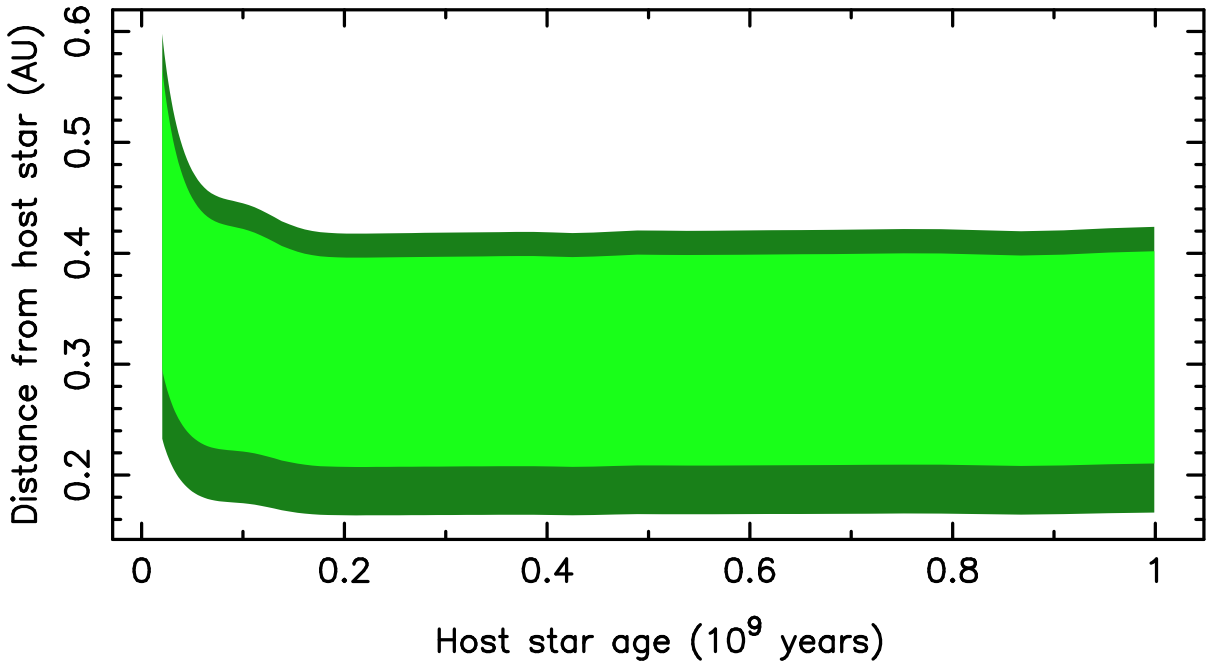}
  \end{center}
  \caption{The evolving HZ for AU~Mic, from the present epoch up to an
    age of 1~Gyr. As per Figure~\ref{fig:hz}, the CHZ and OHZ are
    represented by the light and dark green regions, respectively. The
    HZ regions rapidly contract toward the star during the pre-main
    sequence phase, then start to gradually move away from the star
    after $\sim$200~Myrs.}
  \label{fig:evolve}
\end{figure*}

Using the above described stellar properties, we calculate the extent
of the HZ at the present epoch
\citep{kasting1993a,kane2012a,kopparapu2013b,kopparapu2014}. Specifically,
we adopt the conservative HZ (CHZ) boundaries, defined by the runaway
and maximum greenhouse limits, and the optimistic HZ (OHZ) boundaries,
defined by empirically derived estimates of when Venus and Mars may
have had surface liquid water \citep{kasting1993a,kane2016c}. By
adopting these HZ definitions, we calculated the extent of the CHZ and
OHZ regions to be 0.31--0.60~AU and 0.24--0.63~AU, respectively. A
top-down view of the system, including the orbits of the known
planets, is shown in Figure~\ref{fig:hz}. The star is located at the
center of the crosshairs, and the CHZ and OHZ are represented by the
light green and dark green regions, respectively. Note that the HZ
boundaries are highly dependent on stellar parameter uncertainties
\citep{kane2014a} and distance estimate \citep{kane2018a}. As
described above, the star is both nearby and substantially
characterized, and so the present HZ boundaries are likewise well
defined. However, the star is also young and rapidly evolving, and so
the HZ boundaries are yet to arrive at the locations that will
dominate their main sequence lifetime.


\section{Planetary Habitability Evolution}
\label{evolution}


\subsection{The Evolving Habitable Zone}
\label{evohz}

The evolution of stars on the main sequence, particularly with regards
to their luminosity and effective temperature, results in a subsequent
evolution of the HZ \citep{underwood2003,ramirez2014c}. This HZ
evolution has been studied within the context of stellar masses and
chemical compositions
\citep{young2012a,valle2014b,truitt2015,truitt2017}, and also with
respect to stellar rotation and magnetic activity \citep{gallet2017a}.
The AU~Mic system provides an opportunity to study the evolution of a
planetary system within the context of the early luminosity
environment of the host star as it joins the main sequence. In
particular, the stellar parameters adopted in Section~\ref{system} and
used to calculate the HZ represent a period of rapid luminosity
evolution.

To investigate the effects of the AU~Mic stellar evolution on the
extent of the HZ, we utilize the MESA Isochrones \& Stellar Tracks
(MIST) to calculate the evolutionary track
\citep{paxton2011,paxton2013,paxton2015,choi2016,dotter2016,paxton2018,paxton2019}. Metallicity
measurements for AU~Mic have produced a variety of values, ranging
from $-0.12$~dex \citep{gaidos2014c} to $+0.32$~dex
\citep{neves2013}. We adopt the lower metallicity values, and assume a
metallicity of $-0.1$~dex in the calculation of the evolutionary
track. We further adopt the stellar mass of 0.5~$M_\odot$ (see
Section~\ref{system}) and an initial rotation rate of $v/v_{crit} =
0.0$. This produced stellar properties of $T_\mathrm{eff} = 3712$~K
and $L_\star = 0.082$~$L_\odot$ at an age of 20~Myrs; consistent with
the stellar properties described in Section~\ref{system}.

Figure~\ref{fig:evolve} shows the evolution of the CHZ and OHZ from
the present epoch up to an age of 1~Gyr. The initial HZ boundaries at
an age of 20~Myrs are consistent with those shown in
Figure~\ref{fig:hz}, as expected. The boundaries rapidly shift inward
as the star transitions to the main sequence, reaching their minimum
values at $\sim$200~Myrs, then gradually increase with time. The
period up to 200~Myrs are a particularly important era for the
evolution of terrestrial planets during which the secondary atmosphere
of the planet is degassed from the interior \citep{lammer2008}, a
process that continues well into the stellar main sequence
phase. Furthermore, the presence of two known giant planets interior
to the HZ presents an architecture that may have significant
consequences for terrestrial planets in the HZ. Several dynamical
studies have shown that the formation and migration of giant planets
through the HZ does not necessarily eliminate the presence of
terrestrial planets in that region, and in fact may promote
terrestrial planet formation in the turbulent wake of giant planet
migration \citep{raymond2005b,raymond2006d,fogg2009,bond2010b}. The
location of giant planets also affects the delivery of volatiles to
terrestrial planets and therefore plays a major role in their surface
water inventory \citep{raymond2004a,obrien2018}. Although a dearth of
water is likely to truncate surface habitability, an over-abundance of
water can also directly impede carbonate-silicate cycles and other
processes required for moderating surface environments (see
Section~\ref{evoatmos}). Subsequently, the presence of terrestrial
planets within these evolving HZ regions may provide critical clues
regarding the early processes that can determine the eventual pathway
of potentially habitable surface environments.


\subsection{Early Terrestrial Planetary Atmospheres}
\label{evoatmos}

Young systems, such as AU~Mic, provide potential opportunities to
explore the evolution of the infant stages of planetary
atmospheres. These early stages can play a critical role in
determining the evolutionary pathway that the planetary atmosphere
will follow. For example, the planetary interior, combined with host
star interaction and planetesimal impacts, will drive secondary
atmosphere production and composition
\citep{schlichting2015,kane2020d,kite2020c,oosterloo2021}. The
activity of the host star, particular at young ages, is a major
contributor to atmospheric mass-loss, that both erodes the primary
atmosphere and shapes the evolution of the secondary atmosphere
\citep{zendejas2010,lalitha2018,howe2020}. Moreover, atmospheric
mass-loss during the early stages of planetary evolution can have a
major impact on the subsequent volatile content of the atmosphere and
surface \citep{dong2017b,roettenbacher2017}, which in turn will
influence the overall habitability of the planetary surface
\citep{luger2015b,dong2018a,johnstone2019a}. AU~Mic is a particularly
active star, and has significant variability including significant
spot modulation and frequent white-light flares, both visible in TESS
optical photometry \citep{gilbert2021}. AU Mic~flares span the full
electromagnetic spectrum ranging from x-ray to radio
\citep{leto2000,magee2003}, including EUV flares
\citep{tsikoudi2000b}, which are known to cause photodissociation in
planetary atmospheres.

During subsequent geologic evolution, the early-formed atmosphere can
be largely overprinted by continued degassing, weathering, or loss to
space. As a result, it can be difficult to reconstruct the early
atmospheric state of a terrestrial planet, and learn about the
processes that shaped atmospheric evolution during this time. Young
systems, such as AU Mic, are therefore the best opportunity to study
this important time period in a planet's history. One particularly
uncertain aspect of early terrestrial planet atmospheres is whether
the volatiles the planet acquires during formation (especially H \& C)
will mostly be outgassed to the atmosphere during, or just after, the
formation process, or remain in the interior. For HZ planets with
liquid surface water, subsequent weathering processes drive
atmospheric CO$_2$ concentration, and hence climate, to a state
dictated by the balance between the CO$_2$ outgassing and silicate
weathering rates \citep[e.g.][]{walker1981,berner1983b,berner1997}; at
this point the initial atmospheric composition, at least for CO$_2$,
is lost \citep[e.g.][]{driscoll2013,foley2015,foley2019}. However, it
can take up to $\sim$1~Gyr for the initial atmospheric state to be
lost due to later degassing and weathering
\citep[e.g.][]{driscoll2013,foley2015,foley2019}. Therefore, a planet
sitting in the HZ in a system like AU~Mic, that is $\sim$10s of Myrs
old, would likely still have an atmosphere in transition from it's
initial state, unless atmospheric escape, or any other loss process,
is fast enough to remove an atmosphere outgassed during accretion or
the magma ocean phase in less than $\sim 10$ Myrs.

To expand on the timescale for a terrestrial planet's initial
atmospheric state to be overprinted by weathering and outgassing, we
use a coupled model of the carbonate-silicate cycle and interior
thermal evolution. According to our calculation of the HZ for AU Mic,
there is a range of orbital distances between $\approx 0.3-0.4$ AU
where a planet would remain in the HZ throughout the star's evolution,
including during the pre-main sequence phase. We therefore assume the
planets in our models are always in the HZ and have liquid water
present on their surfaces. The models further assume an Earth-like
water inventory. Planets with moderately higher water inventories,
such that oceans cover the surface to depths of $\sim 10$ km or less,
can likely still support a carbonate-silicate cycle
\citep[e.g.][]{abbot2012a,hayworth2020b}. However, planets with much
larger water inventories may form high pressure ice layers that
prevent silicate weathering
\citep{levi2017a,glaser2020a,krissansentotton2021b}. We also neglect
atmospheric loss in our main set of models, but explore how
atmospheric loss would change our results in
Section~\ref{otherfactors}.

We calculate how long it takes for hypothetical
terrestrial planets to reach approximately the same climate state when
started from an initially hot, CO$_2$-rich atmosphere versus an
initially cold, CO$_2$-poor atmosphere. The model is from
\citet{foley2019}, and calculates rates of CO$_2$ outgassing,
weathering, and atmospheric CO$_2$ concentrations for rocky planets in
a stagnant-lid regime, as the typical regime of tectonics for rocky
planets is unknown and difficult to predict
\citep[e.g.][]{valencia2007c,oneill2007d,korenaga2010b,vanheck2011,foley2012,lenardic2012,noack2014b}. The
model calculates the rate of both volcanic degassing of mantle C and
metamorphic degassing of crustal C, that is released as crust is
buried to high temperature-pressure conditions. The model assumes
seafloor weathering is the dominant weathering process, as forming
large subaerial continents may be less likely on stagnant-lid
planets. The seafloor weathering formulation is after
\citet{krissansentotton2017}.

We ran a set of $10^4$ models for an Earth-like planet (same mass and
core-mass fraction as Earth), randomly sampling from uncertainty
ranges of the following key model parameters: the mantle reference
viscosity, $\mu_{\mathrm{ref}}$, which is the mantle viscosity at
Earth's present day mantle potential temperature of 1350$^{\circ}$C;
the initial mantle potential temperature, $T_{\mathrm{init}}$; the
initial radiogenic heat production rate in the mantle, $Q_0$; the
total budget of CO$_2$ in the mantle and surface reservoirs,
$C_{\mathrm{tot}}$; the reference seafloor weathering rate,
$F_{\mathrm{sfw}}^{*}$, which is defined as Earth's estimated present
day value; the activation energy for seafloor weathering,
$E_{\mathrm{sfw}}$; and the exponent governing the dependence of the
seafloor weathering rate on the rate of extrusive volcanism,
$\beta$. For each model we randomly draw values of the above
parameters from uniform distributions, then run one model where all of
the planet's CO$_2$ initially resides in the atmosphere, and one where
the initial surface temperature is set to 273~K, such that most of the
CO$_2$ initially resides in the mantle. The amount of CO$_2$ that
resides in the mantle initially is a complex function of the planet's
initial bulk C budget and the pressure, temperature, and oxygen
fugacity of equilibration during the formation of the central Fe-core
\cite[][ and references therein]{kane2020a}. We then calculate how
long it takes for these climate states to converge, in terms of the
total planet age after system birth (the convergence age), and the
time elapsed after degassing first begins (the convergence
time). Stellar irradiation is assumed to be the same as what the
modern Earth receives, and is held fixed in the models. The effect of
varying stellar irradiation is discussed below.

\begin{figure*}
  \begin{center}
    \includegraphics[angle=270,width=16.0cm]{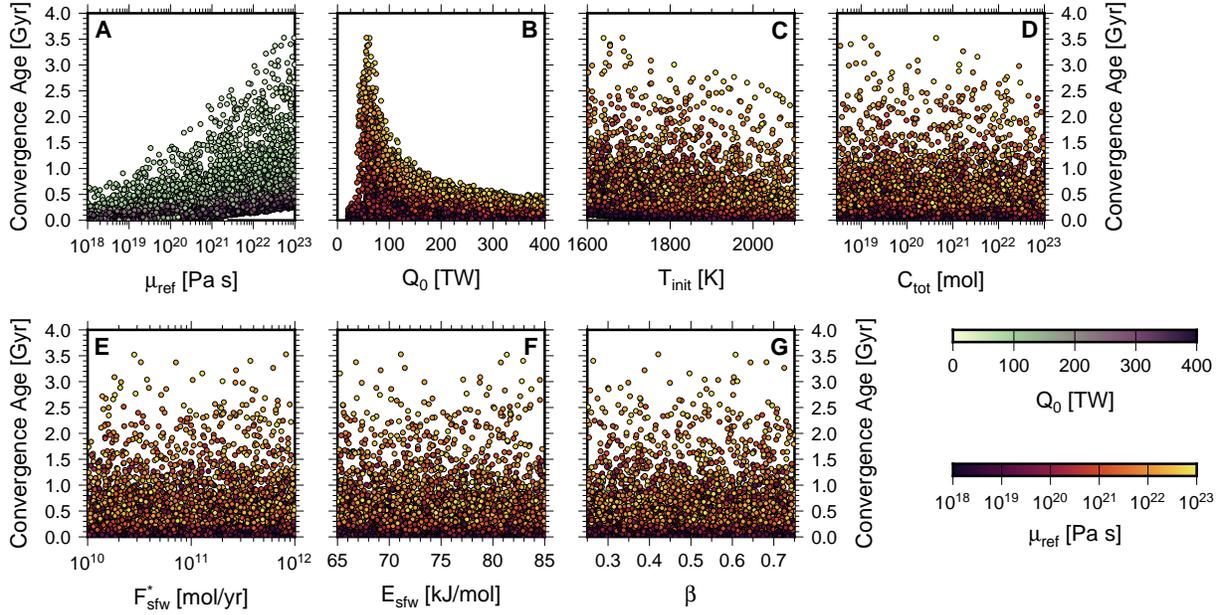}
  \end{center}
  \caption{Planet age, measured after system formation, when an
    initially hot, CO$_2$-rich climate and initially cold, CO$_2$-poor
    climate states converge due to the negative feedbacks of the
    carbonate-silicate cycle. Convergence age is shown as a function
    of mantle reference viscosity (A), initial radiogenic heat
    production rate in the mantle (B), initial mantle potential
    temperature (C), total carbon budget of the mantle and surface
    reservoirs (D), reference seafloor weathering rate (E), activation
    energy for seafloor weathering (F), and dependence parameter of
    seafloor weathering on extrusive volcanism rate (G). Each symbol
    plotted represents one pair of model runs. Symbols are colored
    either by mantle reference viscosity or initial heat production
    rate, as denoted by the colorbars. }
  \label{fig:convage}
\end{figure*}

\begin{figure*}
  \begin{center}
    \includegraphics[angle=270,width=16.0cm]{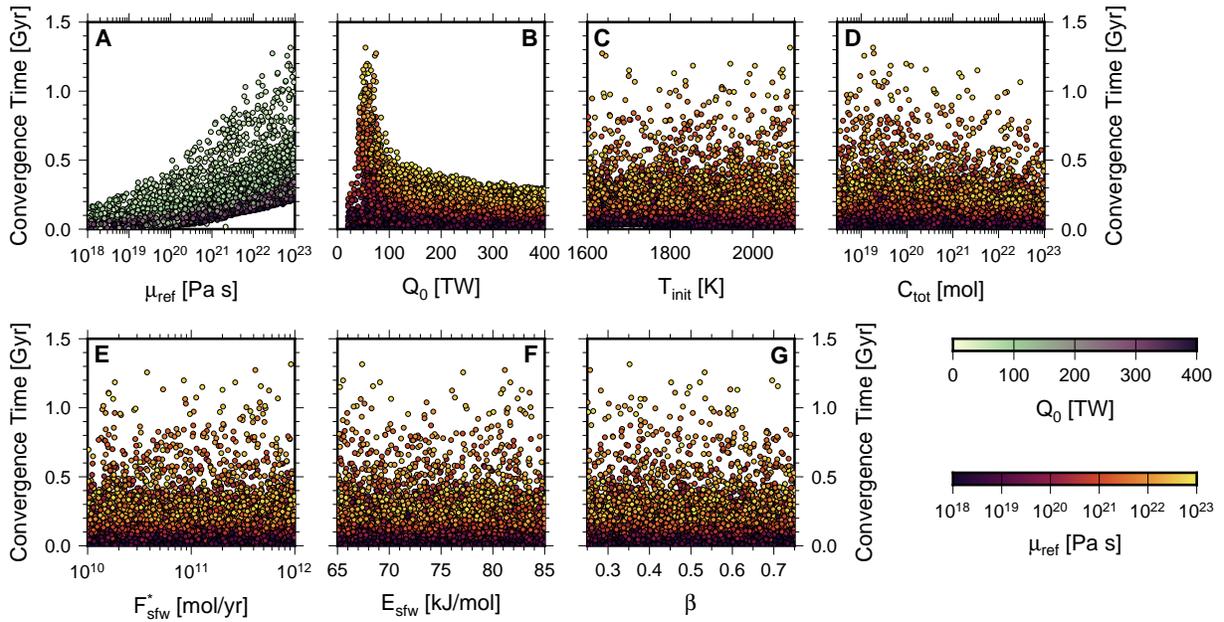}
  \end{center}
  \caption{Same as Figure \ref{fig:convage}, except climate
    convergence is measured as the time elapsed after volcanism and
    degassing has begun.}
  \label{fig:convtime}
\end{figure*}

We find that the mantle reference viscosity and the initial heat
production rate have the biggest influence on both the climate
convergence time and age (Figures \ref{fig:convage} \&
\ref{fig:convtime}). The reason for this is that the ability to
quickly outgas CO$_2$ from the mantle is the primary control on how
long it takes the two different initial climate states to
converge. Silicate weathering can adjust to balance the degassing rate
in $\sim 10^5-10^6$ years \citep{berner1997}, so a planet starting
with an initially hot, CO$_2$-rich climate rapidly reaches a state
where weathering and degassing are approximately in balance. However,
degassing is much slower, so a much longer time is needed for a planet
starting from an initially cold, CO$_2$-poor atmosphere to warm
up. The climate convergence time and age are therefore both primarily
controlled by how quickly degassing can build up CO$_2$ on a planet
where the atmosphere is initially CO$_2$-poor. High rates of heat
production and low reference viscosities favor vigorous mantle
convection and rapid early outgassing, and hence lead to low
convergence times and ages (Figures \ref{fig:convage}A-B \&
\ref{fig:convtime}A-B). There is also a peak in convergence time and
age at $Q_0 \approx 70$ TW. Decreasing initial heat production rate
even further leads to many planets where outgassing never occurs, as
the interiors never became hot enough to melt. Such planets are not
plotted here, as initial atmospheric states can not be erased by the
carbonate-silicate cycle on such planets when there is no interior
outgassing. The only planets able to outgas with low $Q_0 < \approx
70$ TW are those with low reference viscosities and hence more
vigorous convection.

The parameters governing seafloor weathering and the total CO$_2$
budget have a much smaller effect (Figures \ref{fig:convage}D-G \&
\ref{fig:convtime}D-G). Both climate convergence times and ages are
largely insensitive to these parameters, other than a slight
dependence of the convergence time on $C_{\mathrm{tot}}$. These
results are consistent with the argument above that the primary factor
controlling climate convergence is degassing from the
interior. Silicate weathering always acts faster than degassing, so
varying the parameters for silicate weathering has little effect on
climate convergence. The time when climate states converge after
degassing starts weakly declines with increasing $C_{\mathrm{tot}}$,
as more carbon in the system leads to higher degassing rates, and
hence a faster convergence of the initially cold climate to the
initially hot climate state.  

Generally the same trends are seen when
looking at either the time when climate states converge after
degassing has begun, or the planet age when climate states
converge. The convergence ages are higher, reaching
$\approx$3.5--4~Gyrs, while convergence times only reach
$\approx$1.3--1.4~Gyrs. It can take upwards of 2--2.5~Gyrs for
degassing to begin in some cases, leading to this difference. In
particular, the same characteristics that lead to a longer convergence
time, by keeping degassing rates low, can also delay the beginning of
degassing entirely: a high reference viscosity or low heat production
rate. In addition, a low initial mantle temperature also delays
degassing, as the mantle must heat up before it can melt in this
case. As a result, convergence age shows some dependence on
$T_{\mathrm{init}}$, with lower $T_{\mathrm{init}}$ leading to larger
convergence ages (Figure \ref{fig:convage}C). However, convergence
time shows no discernible dependence on $T_{\mathrm{init}}$ (Figure
\ref{fig:convtime}C), because once degassing has begun, the influence
of the initial mantle temperature has already been lost. The rate of
degassing after this point will solely depend on the vigor of
convection, which is controlled by the mantle viscosity and heating
power available to drive convection. The convergence time depends
modestly on the total carbon budget, as described above, but the
convergence age does not (Figure \ref{fig:convage}D). The planet age
when degassing begins does not depend on $C_{\mathrm{tot}}$, as long
as there are at least some interior volatiles available to degas,
because it only depends on how long it takes for the mantle to heat up
enough to melt (neglecting any possible dependence of the mantle
solidus on mantle C content). The way a higher carbon budget helps
enhance degassing rates once volcanism has begun is swamped out by the
time it takes for volcanism to begin when looking convergence age.

\begin{figure*}
  \begin{center}
    \includegraphics[angle=270,width=16.0cm]{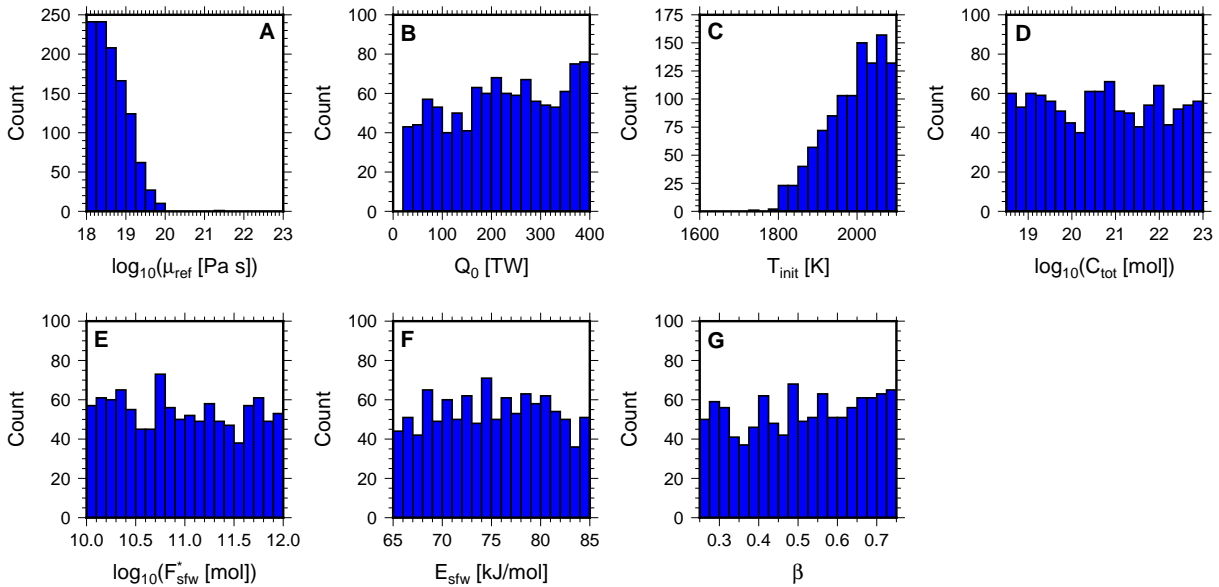}
  \end{center}
  \caption{Histograms of model parameters for all models that produce
    climate convergence ages less the 20 Myrs, or approximately the
    age of the AU Mic system. Shown are distributions of mantle
    reference viscosity (A), initial heat production rate (B), initial
    mantle temperature (C), total carbon budget (D), reference
    seafloor weathering rate (E), activation energy for seafloor
    weathering (F), and dependence parameter of seafloor weathering on
    extrusive volcanism rate (G).}
  \label{fig:conv20myrs}
\end{figure*}

Overall, $\approx 90$ \% of our models produce climate convergence
ages that are greater than the age of the AU Mic system. Habitable
terrestrial exoplanets in systems of similar age to AU Mic are
therefore highly likely to have atmospheres that are still in some
state of transition from their initial conditions, and therefore
inform the range of initial atmospheric conditions on terrestrial
exoplanets. Models that produce convergence ages $< 20$~Myrs are
preferentially those with low reference viscosities $< 10^{20}$ Pa s,
and high initial mantle temperatures $> 1800$~K
(Figure~\ref{fig:conv20myrs}). Very young convergence ages are seen
for the full range of initial radiogenic heat production rates we
explored, so there is no strong connection between heat budget and
producing convergence ages $< 20$ Myrs. For a very young convergence
age, it is important for extensive volcanism to begin very early in a
planet's history, and this is more readily accomplished through an
initially hot mantle than through high rates of heat production. Even
with a high internal heating rate, there would still be a significant
time lag before volcanism can begin if the initial mantle temperature
is low. Consistent with our other results, neither the total C budget
nor the parameters for weathering exert a significant control on
producing very short convergence ages. In sum, only planets with high
initial mantle temperatures and low reference viscosities would be
expected to have lost the influence of their initial atmospheric state
by the $\approx$20~Myr age of AU~Mic, absent other factors, such as
atmospheric escape, we discuss next.


\subsection{Additional Atmosphere Evolution Factors}
\label{otherfactors}

Our models of outgassing, the carbonate-silicate cycle, and climate
evolution ignore stellar evolution, when in reality stars are
typically rapidly evolving during this early phase of their
lifetimes. In the case of AU~Mic, luminosity is decreasing as the star
evolves towards the main sequence. This shift in luminosity changes
the position of the HZ, but not by so much that planets can not stay
within the HZ throughout the entire evolution. We therefore first
discuss how changing solar luminosity would change our model results
for a planet that always remains in the HZ. More extreme evolution,
where planets may start in a runaway greenhouse state before ending up
in the HZ after the star reaches the main sequence, will be discussed
later. For a planet that is always in the HZ during the star's
pre-main sequence phase, decreasing luminosity would lead to a cooling
climate. The carbonate-silicate cycle on such a planet would then
respond by acting to boost the atmospheric CO$_2$ content, as a result
of a temporary slowdown in weathering rates caused by the cooling
climate. A background secular trend that leads to climate cooling, and
hence a buildup of atmospheric CO$_2$ through the carbonate-silicate
cycle, would act to increase the climate convergence times and ages we
presented in Section~\ref{evoatmos}, because it would require even
more CO$_2$ to be outgassed from the mantle in the initially cold
climate state, before initial conditions are erased.

For increasing stellar luminosity, the effect runs the other
way. Higher luminosity favors less CO$_2$ in the atmosphere due to
carbonate-silicate cycle feedbacks. Less CO$_2$ from the interior
needs to be outgassed to warm an initially cool climate, and therefore
climate convergence ages and times would be shorter. These effects can
of course be superimposed in the same system, as the star's luminosity
decreases during pre-main sequence evolution, and then increases when
it reaches the main sequence. Given the timescales of stellar
evolution, low mass stars would favor longer preservation of initial
climate states, while high mass stars would favor these states being
erased more quickly, as long as the planet remains in the HZ the whole
time and escape rates are low enough for the atmosphere to be
retained.

M-dwarf stars, especially the lowest mass M-dwarf stars, present a
more extreme case, as they remain in a superluminous state during
their early life for substantially longer than G-dwarfs, like the Sun
\citep{luger2015b}. In such systems, the HZ boundaries can shift
significantly, such that any planets in the HZ when the star reaches
the main sequence will have been in a runaway greenhouse state during
the pre-main sequence phase. In a runaway greenhouse climate, the lack
of liquid water will largely eliminate silicate weathering, meaning
all outgassed CO$_2$ will remain in the atmosphere. In addition, rapid
escape of both H and heavier species is likely
\citep[e.g.][]{ramirez2014c,luger2015b}. Such rapid escape could
quickly remove both any initial atmosphere left after magma ocean
solidification, and any atmosphere outgassed by subsequent
volcanism. Young planets in such a situation may still show an
atmosphere in a transitional phase, but in this case a phase where
rapid escape is depleting the atmosphere. If enough volatiles remain
in the planetary interior after the star settles on to the main
sequence, an atmosphere may be able to re-form by volcanic outgassing.

\begin{figure*}
  \begin{center}
    \includegraphics[angle=270,width=16.0cm]{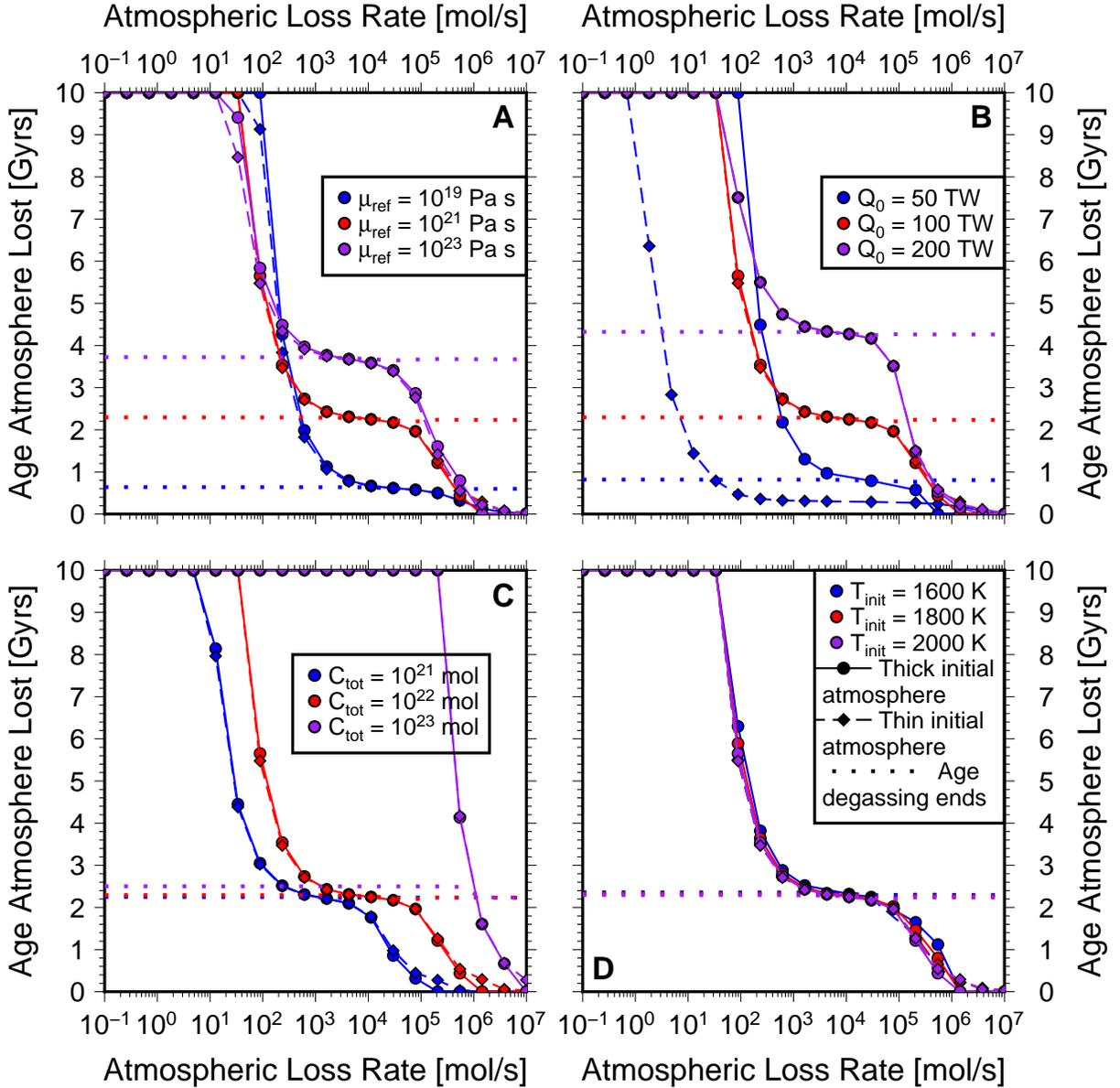}
  \end{center}
  \caption{Planet age when the atmosphere is completely lost as a
    function of an imposed, constant atmospheric loss rate for varying
    mantle reference viscosity, $\mu_{\mathrm{ref}}$ (A), initial heat
    production rate, $Q_0$ (B), mantle carbon budget,
    $C_{\mathrm{tot}}$ (C), and initial mantle temperature,
    $T_{\mathrm{init}}$ (D). All models assume $F_{\mathrm{sfw}}^{*} =
    5 \times 10^{11}$ mol$\cdot$yr$^{-1}$, $E_{\mathrm{sfw}} = 75$
    kJ$\cdot$mol$^{-1}$, and $\beta = 0.5$. Unless otherwise
    specificed, models also assume $\mu_{\mathrm{ref}}=10^{21}$
    Pa$\cdot$s, $Q_0 = 100$ TW, $C_{\mathrm{tot}} = 10^{22}$ mol, and
    $T_{\mathrm{init}} = 2000$ K. Model results are color coded based
    on the legends in each panel. Circles connected by solid lines are
    used for models started with a CO$_2$-rich initial atmosphere, and
    diamonds connected by dashed lines denote models with an initially
    CO$_2$-poor atmosphere. In many models sets there is litte
    difference between the two initial conditions, so symbols lie on
    top of each other. Dotted lines show the age when interior
    degassing stops, color coded based on the legend for each
    panel. Models are run for 10 Gyrs, so models where an atmosphere
    is retained after 10 Gyrs are considered to have not lost their
    atmospheres.}
  \label{fig:atm_loss1}
\end{figure*}

\begin{figure*}
  \begin{center}
    \includegraphics[angle=270,width=16.0cm]{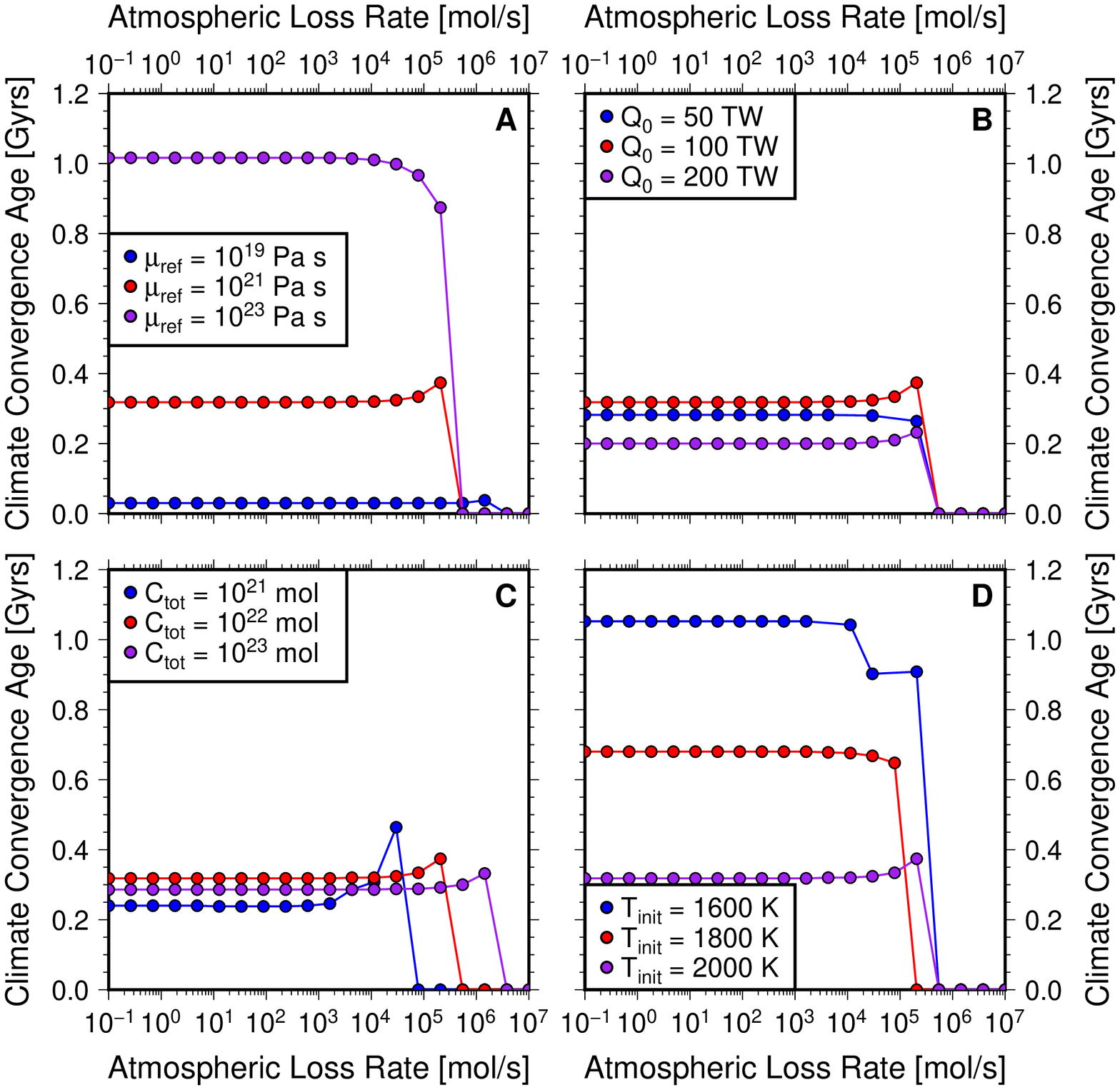}
  \end{center}
  \caption{Climate convergence age as a function of an imposed,
    constant atmospheric loss rate for varying mantle reference
    viscosity, $\mu_{\mathrm{ref}}$ (A), initial heat production rate,
    $Q_0$ (B), mantle carbon budget, $C_{\mathrm{tot}}$ (C), and
    initial mantle temperature, $T_{\mathrm{init}}$ (D). Other model
    parameters are the same as listed in the caption to Figure
    \ref{fig:atm_loss1}. Climate convergence age drops to zero when
    atmospheric loss is rapid enough to prevent any appreciable
    atmosphere from forming. }
  \label{fig:atm_loss2}
\end{figure*}

To quantify how atmospheric loss processes, either driven by
hydrodynamic escape or non-thermal erosion, influence our CO$_2$
outgassing and atmosphere evolution model results, we added a constant
atmospheric loss rate to our models and varied this loss rate over a
wide range (Figure \ref{fig:atm_loss1} \& \ref{fig:atm_loss2}). We ran
sets of models individually varying the parameters found to influence
climate convergence age the most: mantle reference viscosity, initial
internal heating rate, total carbon budget, and initial mantle
temperature. Our thermal evolution models are tracking CO$_2$
outgassing from the interior, so we assume a CO$_2$ dominated
atmosphere. Our imposed atmospheric loss rates are therefore rates of
CO$_2$ loss in mol$\cdot$s$^{-1}$.

At low atmospheric loss rates, the atmosphere is retained for at least
10~Gyrs (the ending time of our models), even though volcanism and
degassing ends far earlier (Figure~\ref{fig:atm_loss1}). The latest
that volcanism lasts in any of our models shown in Figure
\ref{fig:atm_loss1} is $\approx 4.5$ Gyrs. After degassing has ended,
the atmosphere is simply lost at the imposed loss rate, until it is
entirely depleted. Atmospheric retention time therefore scales
inversely with loss rate. As a result, there is a threshold loss rate
where the atmosphere can not be retained for 10 Gyrs or longer, and
beyond this threshold the planet age where the atmosphere is totally
lost decreases linearly with increasing loss rate. The behavior
changes then when loss rates are high enough to entirely remove the
atmosphere before degassing has ceased. Total atmospheric loss can
happen while degassing is still active when the loss rate exceeds the
degassing rate; in this case the existing atmosphere can be lost and
degassing is unable to replenish the atmosphere. All degassed
volatiles are rapidly lost to space. Degassing rate decreases over
time as the planet cools, before falling to zero when volcanism
stops. As a result, large atmospheric loss rates are needed to push
the age when the atmosphere is lost further back in time, to younger
planet ages. Finally, the planet age when the atmosphere is lost drops
towards zero when the loss rate exceeds even the peak degassing rate
the planet experiences during its evolution. For loss rates this high
the planet can effectively never form an atmosphere; anything degassed
from the interior is rapidly lost to space.

The planet age where degassing ends is a function of mantle reference
viscosity and initial heat budget \citep[e.g.][]{foley2018a}, leading
to a corresponding change in the age where the atmosphere is lost. For
atmospheric loss rates high enough to cause total atmosphere loss
before degassing ends, but low enough to still allow an atmosphere to
form, higher reference viscosities and higher initial heat production
rates lead to longer retention of an atmosphere, because these effects
prolong volcanism and degassing (Figure~\ref{fig:atm_loss1}A \&
B). When the mantle carbon budget is increased, interior degassing
rates are higher and a thicker atmosphere can form. As a result, the
larger the carbon budget, the longer the atmosphere can be retained
for a given loss rate (Figure \ref{fig:atm_loss1}C). Initial mantle
temperature does not significantly influence rates of degassing over
time, or when degassing ends, so it does not significantly influence
the planet age where the atmosphere is lost. Finally, in most cases
the initial atmospheric state does not affect the age when the
atmosphere is lost. The one exception is the case with low initial
heat production ($Q_0 = 50$ TW, Figure \ref{fig:atm_loss1}B). In this
case degassing rates are low and end early in the planet's lifetime,
after less than 1 Gyr. As a result, the initially CO$_2$-poor
atmosphere always stays thinner than the initially CO$_2$-rich
atmosphere, and hence is lost more quickly. Furthermore, as explained
above, with initial heating rates this low planets with larger
reference viscosities would not experience outgassing at all,
especially with low initial mantle temperatures.

When atmospheric loss rates are high enough that the planet
essentially never forms an atmosphere, or any primordial atmosphere is
lost on a timescale of less than $10^6-10^7$ yrs, then the issue of
how long it takes two initially different climate states to converge
is no longer relevant. The atmosphere is rapidly lost and observers
would find a barren rock world. However, at lower loss rates that
still allow an atmosphere to form, atmospheric loss could potentially
influence the climate convergence ages and times presented in
Section~\ref{evoatmos}. On the contrary, we find that the climate
convergence age is not strongly influenced by atmospheric loss, for
loss rates low enough to allow an atmosphere to form (Figure
\ref{fig:atm_loss2}). There is some effect for loss rates approaching
the transition point where loss processes would prevent atmosphere
formation entirely. The climate convergence age can either decrease or
increase by $\sim 10-100$ Myrs depending on whether atmospheric loss
acts more to remove an initially thick atmosphere or slow the buildup
of an atmosphere by degassing when starting with an initially thin
atmosphere. However, the largest effects of atmospheric loss on
climate convergence age are seen for planets where the convergence age
is already long, $\sim 0.1-1$ Gyrs. For these planets, the effects of
atmospheric loss are not sufficient to drive the convergence age below
the age of the AU Mic system, except in the case of rates high enough
to entirely remove the atmosphere.

Hydrodynamic escape driven by stellar extreme ultraviolet radiation is
likely the dominant loss process for very young terrestrial planets
\citep[e.g.][]{ramirez2014c,luger2015b}. Stellar extreme ultraviolet
fluxes, $S_{\mathrm{EUV}}$, can be well fit with simple
parameterizations based on stellar type
\citep[e.g.][]{ribas2005}. $S_{\mathrm{EUV}}$ fluxes decrease over
time, so rates of hydrodynamic escape will decline as well. Thus, in
addition to constraining the atmospheric loss rates planets around
young stars might experience, the duration of these high loss rates is
also critical. Rapid early escape could remove any primordial
atmosphere and prevent buildup of an outgassed atmosphere, but as
these escape rates decline continued volcanism could then eventually
produce a secondary atmosphere. Moreover, determining the escape rate
a particular planet would experience is challenging, as this requires
knowledge of the mixing ratios of the gases making up the atmosphere
as a function of height. If hydrogen is abundant in the upper
atmosphere, then escape is energy limited, which represents the upper
bound on escape rates. Rapid hydrogen escape can also drag away
heavier species, such as O or CO$_2$. However, if the H mixing ratio
is low, escape will be diffusion limited and much slower. Determining
mixing ratios of different atmospheric species is beyond the scope of
our simple outgassing models, but we can use previous studies on water
loss as a first order guide.

\cite{ramirez2014c} \& \cite{luger2015b} show that H escape is rapid
for planets in the main sequence habitable zone around low mass
M-dwarf stars, due to the long pre-main sequence phase for these stars
and the close-in main sequence habitable zone. The resulting flows of
escaping H driven by high $S_{\mathrm{EUV}}$ fluxes are also capable
of dragging heavier species, included CO$_2$, with
them. \cite{ramirez2014c} find that the equivalent of $\approx
200-4000$ times the molar quantity of H stored in Earth's oceans can
be lost for planets around the smallest M8 stars over the course of
their $\sim 1$ Gyr long pre-main sequence phase. This corresponds to
loss rates of $\sim 10^8-10^{10}$ mol$\cdot$s$^{-1}$ of H$_2$. Such
rapidly escaping H would also drag away substantial quantities of
CO$_2$, though with CO$_2$ escaping rates of H escape would be
lower. Still, the lowest mass M dwarf stars can likely have nearly
their entire volatile abundances stripped away by pre-main sequence
$S_{\mathrm{EUV}}$ fluxes. Even continued volcanism after the star
reaches the main sequence may not be able to replenish the atmosphere,
as rapid early outgassing when the interior is hot can deplete the
mantle of volatiles within $\sim 1$ Gyr in our models.

Rather than being in a state of transition where the
carbonate-silicate cycle is acting to bring the initial climate into
an approximate steady-state between weathering and outgassing fluxes,
planets around low mass M-dwarfs may instead have their volatiles
entirely stripped. Observations of such planets in young systems are
thus more likely to show runaway greenhouse climates and atmospheres
being rapidly removed. Even if such planets are able to re-form an
atmosphere after their host star reaches the main sequence, the
initial atmospheric state after planet formation and any potential
magma ocean solidification will have been lost. Planets re-forming
their atmospheres will all be starting from similar states:
essentially barren rocky worlds replenishing the atmosphere with
volcanic gases.

More massive stars, including M1 stars and FGK stars, are more likely
to retain atmospheres. Escape rates during these systems' early
evolution can still be large, e.g. $\sim 10^6-10^8$ mol$\cdot$s$^{-1}$
of H$_2$ for planets around M1 stars, which can lose 0.5-25 Earth
ocean masses of H$_2$ over $\sim 200$ Myrs according to
\cite{ramirez2014c}. However, these rapid escape rates are short
lived, and the timespan of rapid escape shrinks as stellar mass
increases. As a result, volatiles can still be retained after the star
reaches the main sequence. We expect that water and CO$_2$ can
therefore be retained around M1 stars and stars more massive than
this. For M1 stars, early atmospheric escape may still be fast enough,
and persist long enough, to remove an initially thick, CO$_2$-rich
atmosphere formed by outgassing from a magma ocean. In this case
climate convergence may happen sooner than our models in Figures
\ref{fig:convage} \& \ref{fig:convtime} indicate. Planets around young
($< 100$ Myr old) M1 stars may have thin atmospheres due to
atmospheric loss regardless of their initial atmospheric state. A more
substantial atmosphere could then form after the star reaches the main
sequence and escape rates fall. AU Mic itself is an M1 star, so it
falls near this transition point where planets around stars less
massive are likely to have initial atmospheric states erased very
rapidly by escape, while planets in the HZ around more massive stars
are likely to retain their initial atmospheres and have their early
climate evolution governed by the carbonate-silicate cycle feedbacks
modeled in Section~\ref{evoatmos}.

Our geophysical models assume that the planet has fully solidified
from its initial magma ocean phase. This process of magma ocean
solidification occurs over Myr timescales and is limited by the rate
of heat loss at the top of the atmosphere to space. This heat loss to
space must exceed the incident stellar radiation from the host-star
then, for the magma ocean to cool and solidify. \citet{hamano2013}
modeled these competing effects for rocky planets and found that the
lifetime of the magma ocean is dependent on a planet's orbital
distance, which sets the incident radiation flux, and the rate of
water loss at the top of the atmosphere due to hydrodynamic
escape. Their models find that magma oceans can last from a few Myr
for planet's within the HZ of a Sun-like star, to $\sim10^{2}$ Myr for
those interior to it. This is independent of any tidal heating that
may be present, which can potentially expand the magma ocean lifetime
even further \citep{driscoll2015}. At only $\sim20$ Myr old, any rocky
exoplanets orbiting AU Mic may potentially still be in the magma ocean
phase of planetary evolution. Detection of SO$_2$ or alkali-bearing
species NaOH and KOH via atmospheric transmission spectroscopy may
hint at active volcanism on these planets
\citep{kaltenegger2010f,schaefer2012}. These atmospheric species,
however, do not reveal whether this volcanism is from traditional
surface melting or directly from the magma ocean. More work is needed
to understand how we can truly detect the presence of a magma ocean,
however, young planetary systems, such as AU Mic, provide us with an
excellent laboratory for understanding this critical part of planetary
evolution.


\section{Orbital Dynamics}
\label{dynamics}


\subsection{Dynamics of the Known Planets}
\label{known}

The orbital dynamics of young planetary systems play a key role in
determining the long-term architecture of those systems. These early
dynamical effects include interactions with the debris disk and other
planets \citep{raymond2012,desousa2020} and close stellar encounters
in cluster environments
\citep{spurzem2009,kane2018b,vanelteren2019}. Here we explore the
dynamics of the two known planets in the system, adopting the orbital
parameters provided by \citet{cale2021} (see Section~\ref{system}). We
follow the methodology of \citet{kane2019c,kane2021a}, which uses the
Mercury Integrator Package \citep{chambers1999} with a hybrid
symplectic/Bulirsch-Stoer integrator with a Jacobi coordinate system
\citep{wisdom1991,wisdom2006b}. Given the 8.46~day orbital period of
the inner planet, we adopted a time resolution of 0.1~days to ensure
sufficient accuracy during times of closest planet-planet
interactions, as recommended by \citet{duncan1998}.

\begin{figure}
  \includegraphics[angle=270,width=8.5cm]{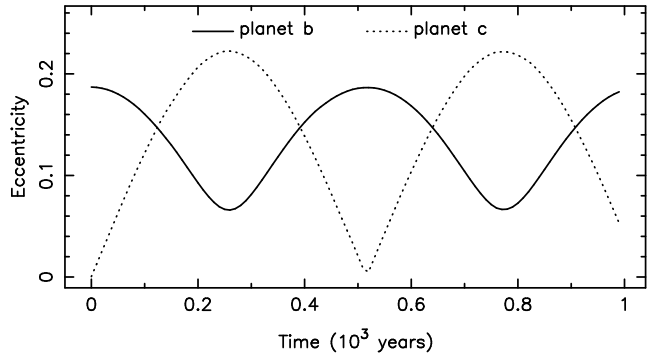}
  \caption{Orbital eccentricity as a function of time for the known
    planets, where planet b and planet c are shown as solid and dotted
    lines, respectively. The planets are long-term stable, but
    exchange significant angular momentum through oscillating
    eccentricities with a period of $\sim$520~yrs.}
  \label{fig:known}
\end{figure}

A suite of dynamical simulations was carried out that explored the
long-term stability of the system for a range of orbital parameters
within the measured uncertainties. The total duration of the
simulations were $10^8$~yrs, equivalent to $\sim$$2\times10^9$ orbits
of planet c. These were all found to be stable over the full
simulation durations. However, the disparity between the measured
orbital eccentricities of the planets creates a high-frequency
exchange of angular momentum. This exchange is visualized in
Figure~\ref{fig:known}, which shows the eccentricity evolution of the
planets over a period of $10^3$~yrs. The period of the eccentricity
variations is $\sim$520~yrs, consistent with the findings of previous
investigations of the time-dependent dynamical effects of non-zero
eccentricities \citep{gladman1993,chambers1996,hadden2018b}. Such
eccentricity oscillations also have the potential to reduce the
stability potential of other planets within the system, including
terrestrial planets within the HZ.


\subsection{Stability Within the Habitable Zone}
\label{hz}

\begin{figure*}
  \begin{center}
    \includegraphics[angle=270,width=16.0cm]{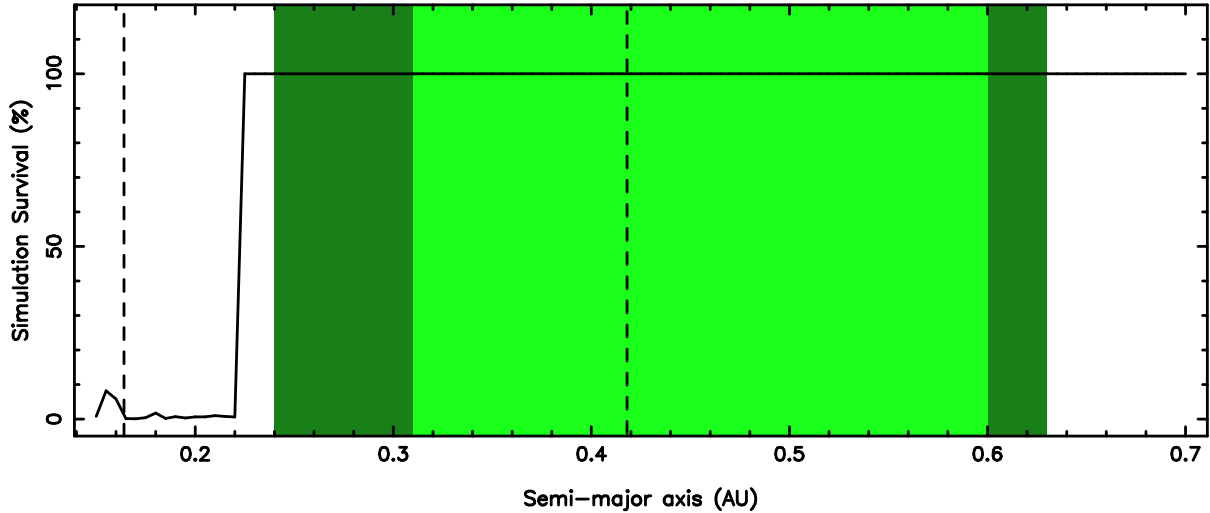}
  \end{center}
  \caption{Plot of the dynamical simulation results for an Earth-mass
    planet injected in the semi-major axis range 0.15--0.70~AU. As for
    Figure~\ref{fig:hz}, the light green and dark green areas
    represent the CHZ and OHZ regions at the present epoch,
    respectively. The vertical dashed lines represent the boundaries
    of the OHZ at a stellar age of 200~Myrs. The solid line shows the
    survival of the Earth-mass planet in terms of the percentage of
    the each simulation for which it remains in the system.}
  \label{fig:simhz}
\end{figure*}

System orbital dynamics, such as those described in
Section~\ref{known}, can also have a dramatic effect on planets within
the HZ, including their long-term stability
\citep{georgakarakos2018,kane2020b} and climate evolution
\citep{way2017a,kane2020e}. Considering the evolution of the HZ
described in Section~\ref{evohz}, we performed a suite of dynamical
simulations that tested the long-term stability of an Earth-mass
planet in initial circular orbits with semi-major axes in the range
0.15--0.70~AU in steps of 0.005~AU. Each simulation was conducted for
$10^6$ years with a time resolution of 0.1~days, as described in
Section~\ref{known}. The outcome of each simulation was assessed based
on whether the injected Earth-mass planet survived or was lost, either
by ejection from the system or lost to the gravitational well of the
host star.

The results of the complete set of simulations are shown in
Figure~\ref{fig:simhz}, where the simulation survival on the vertical
axis refers to the percentage of the simulation for which the injected
planet survived. The light green and dark green areas correspond to
the CHZ and OHZ regions at the present epoch, respectively, equivalent
to those shown in Figure~\ref{fig:hz}. The vertical dashed lines
represent the OHZ boundaries at a stellar age of 200~Myrs, based on
the predicted stellar parameters described in Section~\ref{evohz} and
the HZ boundaries shown in Figure~\ref{fig:evolve}. The simulations
demonstrate that, although a terrestrial planet is able to retain
orbital stability throughout the present HZ, such a planet can
experience severe orbital instability near the inner edge of the main
sequence HZ indicated by the vertical dashed lines. Therefore, a
terrestrial planet that is currently located in the inner half of the
present epoch (pre-main sequence) HZ, will find itself in a long-term
stable orbit in the outer half of the main sequence HZ after the
pre-main sequence phase of the stellar evolution is complete.

\begin{figure}
  \includegraphics[angle=270,width=8.5cm]{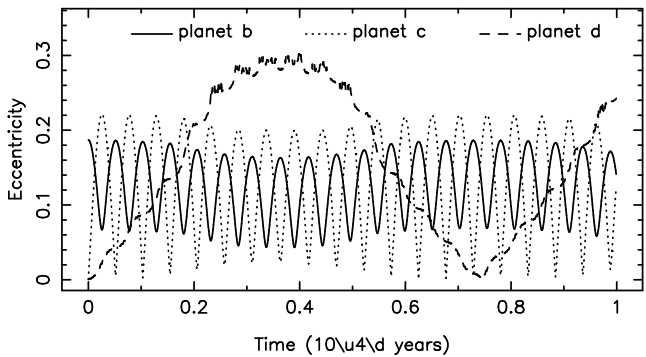}
  \caption{Eccentricity as a function of time for the known planets
    (solid and dotted lines) and a hypothetical terrestrial planet d
    with a semi-major axis of 0.23~AU (dashed line).}
  \label{fig:simplanet}
\end{figure}

Although terrestrial planetary orbits have long-term stable locations
within the inner half of the main sequence HZ, many of these locations
experience significant eccentricity oscillations due to angular
momentum transfer from the inner planets (see Section~\ref{known}).
Figure~\ref{fig:simplanet} shows the eccentricity evolution for the
two known planets (solid and dotted lines) and a simulated terrestrial
planet (planet d) located at 0.23~AU from the host star (dashed
line). This particular location for planet d places the planet at the
very edge of instability, as shown in Figure~\ref{fig:simhz}. In this
case, planet d experiences variations in eccentricity within the range
0.0--0.3, resulting in periods of highly variable insolation flux
received by the planet. Variable eccentricity such as this is known to
play a substantial role in the planetary climate evolution, which can
be dampened depending on the volatile inventory and spin state of the
planet \citep{barnes2013a,linsenmeier2015,way2017a,kane2021a}. In
terms of the early atmospheric evolution described in
Section~\ref{evoatmos}, the variable star--planet separations during
periods of high eccentricity will cause correspondingly variable rates
of atmospheric loss. Additionally, highly eccentric orbital states
will induce tidal heating within the planet which, in turn, will
contribute to the energy budget required for sustained degassing
processes \citep{barnes2009b,driscoll2015}. Thus, the combination of
orbital state, atmospheric loss, and degassing rates creates a complex
interaction between these processes that may serve to have a net
positive or negative effect on long-term habitability, depending on
factors such as the initial volatile inventory.


\section{Conclusions}
\label{conclusions}

The early phase of both star and planet evolution are periods of rapid
change for both classes of objects within the system. The changes
occurring with the star result in a decreasing stellar luminosity,
corresponding to a changing flux environment for the planets within
the system. These changes, in turn, have consequences for the
radiative balance of the new-formed planets, whose interior energy
source will yet be substantial
\citep{wetherill1980a,kleine2002,raymond2004a}. Furthermore, the
dynamical state of young systems can be relatively unstable and
complex compared with their older counterparts, when much of the
significant interactions and tidal dissipation will most likely have
taken place. The AU~mic system, being young, nearby, and containing
two known transiting planets and a debris disk, provides an
opportunity to conduct detailed studies of the early processes of
planet formation and the convergence of potential planetary climates.

In this work, we have described a hypothetical terrestrial planet
within the HZ of the AU~Mic system. The evolutionary processes
regarding the radiation environment and outgassing of the planet,
described in Section~\ref{evolution}, will play a critical role in the
subsequent pathway of the overall planetary surface conditions
\citep{hamano2013,way2020}. Moreover, a non-zero eccentricity induced
by interactions with the other planets in the system will add an
additional time-dependent component to the radiation environment. An
additional effect not considered in our outgassing model is that of
atmospheric erosion, a process that is particularly efficient during
the early stages of star and planet formation
\citep{owen2019a,rodriguezmozos2019,carolan2020}. Continued
observations of AU~Mic are highly encouraged to further characterize
the known planets and potentially detect additional planets in the
system. An improved system architecture model would greatly aid in
determining the viability of stable orbital locations within the
system and how these would evolve through time. Further RV
observations, such as those described by \citet{cale2021}, will
potentially extend the sensitivity of the probed parameter space to
longer periods and smaller masses. Additional {\it TESS} sectors that
observe AU~Mic will likewise refine the properties of the known
planets as well as possibly reveal additional transiting planets
within the system. Direct imaging of the star is lucrative due to its
proximity, and has been used to study the debris disk at millimeter
wavelengths \citep{macgregor2013}, but will likely suffer from zodical
dust contamination at optical wavelengths, inhibiting the detection of
terrestrial planets within the HZ. Regardless, the AU~Mic system is an
incredibly important evolutionary case-study that will serve as a
template for understanding how planets and their atmospheres evolve
during the crucial early stages of their lives.


\section*{Acknowledgements}

This research has made use of the Habitable Zone Gallery at
hzgallery.org. The results reported herein benefited from
collaborations and/or information exchange within NASA's Nexus for
Exoplanet System Science (NExSS) research coordination network
sponsored by NASA's Science Mission Directorate.


\software{Mercury \citep{chambers1999}}



\begin{thebibliography}{}
\expandafter\ifx\csname natexlab\endcsname\relax\def\natexlab#1{#1}\fi
\providecommand{\url}[1]{\href{#1}{#1}}
\providecommand{\dodoi}[1]{doi:~\href{http://doi.org/#1}{\nolinkurl{#1}}}
\providecommand{\doeprint}[1]{\href{http://ascl.net/#1}{\nolinkurl{http://ascl.net/#1}}}
\providecommand{\doarXiv}[1]{\href{https://arxiv.org/abs/#1}{\nolinkurl{https://arxiv.org/abs/#1}}}

\bibitem[{{Abbot} {et~al.}(2012){Abbot}, {Cowan}, \& {Ciesla}}]{abbot2012a}
{Abbot}, D.~S., {Cowan}, N.~B., \& {Ciesla}, F.~J. 2012, \apj, 756, 178,
  \dodoi{10.1088/0004-637X/756/2/178}

\bibitem[{{Baraffe} {et~al.}(2002){Baraffe}, {Chabrier}, {Allard}, \&
  {Hauschildt}}]{baraffe2002}
{Baraffe}, I., {Chabrier}, G., {Allard}, F., \& {Hauschildt}, P.~H. 2002, \aap,
  382, 563, \dodoi{10.1051/0004-6361:20011638}

\bibitem[{{Barnes} {et~al.}(2009){Barnes}, {Jackson}, {Greenberg}, \&
  {Raymond}}]{barnes2009b}
{Barnes}, R., {Jackson}, B., {Greenberg}, R., \& {Raymond}, S.~N. 2009, \apjl,
  700, L30, \dodoi{10.1088/0004-637X/700/1/L30}

\bibitem[{{Barnes} {et~al.}(2013){Barnes}, {Mullins}, {Goldblatt}, {Meadows},
  {Kasting}, \& {Heller}}]{barnes2013a}
{Barnes}, R., {Mullins}, K., {Goldblatt}, C., {et~al.} 2013, Astrobiology, 13,
  225, \dodoi{10.1089/ast.2012.0851}

\bibitem[{{Berner} \& {Caldeira}(1997)}]{berner1997}
{Berner}, R.~A., \& {Caldeira}, K. 1997, Geology, 25, 955,
  \dodoi{10.1130/0091-7613(1997)025<0955:TNFMBA>2.3.CO;2}

\bibitem[{{Berner} {et~al.}(1983){Berner}, {Lasaga}, \&
  {Garrels}}]{berner1983b}
{Berner}, R.~A., {Lasaga}, A.~C., \& {Garrels}, R.~M. 1983, American Journal of
  Science, 283, 641, \dodoi{10.2475/ajs.283.7.641}

\bibitem[{{Bond} {et~al.}(2010){Bond}, {O'Brien}, \& {Lauretta}}]{bond2010b}
{Bond}, J.~C., {O'Brien}, D.~P., \& {Lauretta}, D.~S. 2010, \apj, 715, 1050,
  \dodoi{10.1088/0004-637X/715/2/1050}

\bibitem[{{Cale} {et~al.}(2019){Cale}, {Plavchan}, {LeBrun}, {Gagn{\'e}},
  {Gao}, {Tanner}, {Beichman}, {Xuesong Wang}, {Gaidos}, {Teske}, {Ciardi},
  {Vasisht}, {Kane}, \& {von Braun}}]{cale2019}
{Cale}, B., {Plavchan}, P., {LeBrun}, D., {et~al.} 2019, \aj, 158, 170,
  \dodoi{10.3847/1538-3881/ab3b0f}

\bibitem[{{Cale} {et~al.}(2021){Cale}, {Reefe}, {Plavchan}, {Tanner}, {Gaidos},
  {Gagn{\'e}}, {Gao}, {Kane}, {B{\'e}jar}, {Lodieu}, {Anglada-Escud{\'e}},
  {Ribas}, {Pall{\'e}}, {Quirrenbach}, {Amado}, {Reiners}, {Caballero}, {Rosa
  Zapatero Osorio}, {Dreizler}, {Howard}, {Fulton}, {Xuesong Wang}, {Collins},
  {El Mufti}, {Wittrock}, {Gilbert}, {Barclay}, {Klein}, {Martioli},
  {Wittenmyer}, {Wright}, {Addison}, {Hirano}, {Tamura}, {Kotani}, {Narita},
  {Vermilion}, {Lee}, {Geneser}, {Teske}, {Quinn}, {Latham}, {Esquerdo},
  {Calkins}, {Berlind}, {Zohrabi}, {Stibbards}, {Kotnana}, {Jenkins},
  {Twicken}, {Henze}, {Kidwell}, {Burke},
  {Villase\{\textbackslash\raisebox{-0.5ex}\textasciitilde n\}or}, \&
  {Boyd}}]{cale2021}
{Cale}, B., {Reefe}, M., {Plavchan}, P., {et~al.} 2021, arXiv e-prints,
  arXiv:2109.13996.
\newblock \doarXiv{2109.13996}

\bibitem[{{Carolan} {et~al.}(2020){Carolan}, {Vidotto}, {Plavchan}, {Villarreal
  D'Angelo}, \& {Hazra}}]{carolan2020}
{Carolan}, S., {Vidotto}, A.~A., {Plavchan}, P., {Villarreal D'Angelo}, C., \&
  {Hazra}, G. 2020, \mnras, 498, L53, \dodoi{10.1093/mnrasl/slaa127}

\bibitem[{{Chambers}(1999)}]{chambers1999}
{Chambers}, J.~E. 1999, \mnras, 304, 793,
  \dodoi{10.1046/j.1365-8711.1999.02379.x}

\bibitem[{{Chambers} {et~al.}(1996){Chambers}, {Wetherill}, \&
  {Boss}}]{chambers1996}
{Chambers}, J.~E., {Wetherill}, G.~W., \& {Boss}, A.~P. 1996, \icarus, 119,
  261, \dodoi{10.1006/icar.1996.0019}

\bibitem[{{Choi} {et~al.}(2016){Choi}, {Dotter}, {Conroy}, {Cantiello},
  {Paxton}, \& {Johnson}}]{choi2016}
{Choi}, J., {Dotter}, A., {Conroy}, C., {et~al.} 2016, \apj, 823, 102,
  \dodoi{10.3847/0004-637X/823/2/102}

\bibitem[{{David} {et~al.}(2019{\natexlab{a}}){David}, {Petigura}, {Luger},
  {Foreman-Mackey}, {Livingston}, {Mamajek}, \& {Hillenbrand}}]{david2019c}
{David}, T.~J., {Petigura}, E.~A., {Luger}, R., {et~al.} 2019{\natexlab{a}},
  \apjl, 885, L12, \dodoi{10.3847/2041-8213/ab4c99}

\bibitem[{{David} {et~al.}(2019{\natexlab{b}}){David}, {Cody}, {Hedges},
  {Mamajek}, {Hillenbrand}, {Ciardi}, {Beichman}, {Petigura}, {Fulton},
  {Isaacson}, {Howard}, {Gagn{\'e}}, {Saunders}, {Rebull}, {Stauffer},
  {Vasisht}, \& {Hinkley}}]{david2019b}
{David}, T.~J., {Cody}, A.~M., {Hedges}, C.~L., {et~al.} 2019{\natexlab{b}},
  \aj, 158, 79, \dodoi{10.3847/1538-3881/ab290f}

\bibitem[{{Davies}(1980)}]{davies1980}
{Davies}, D.~W. 1980, \icarus, 42, 145, \dodoi{10.1016/0019-1035(80)90252-3}

\bibitem[{{de Sousa} {et~al.}(2020){de Sousa}, {Morbidelli}, {Raymond},
  {Izidoro}, {Gomes}, \& {Vieira Neto}}]{desousa2020}
{de Sousa}, R.~R., {Morbidelli}, A., {Raymond}, S.~N., {et~al.} 2020, \icarus,
  339, 113605, \dodoi{10.1016/j.icarus.2019.113605}

\bibitem[{{Dong} {et~al.}(2017){Dong}, {Huang}, {Lingam}, {T{\'o}th},
  {Gombosi}, \& {Bhattacharjee}}]{dong2017b}
{Dong}, C., {Huang}, Z., {Lingam}, M., {et~al.} 2017, \apjl, 847, L4,
  \dodoi{10.3847/2041-8213/aa8a60}

\bibitem[{{Dong} {et~al.}(2018){Dong}, {Jin}, {Lingam}, {Airapetian}, {Ma}, \&
  {van der Holst}}]{dong2018a}
{Dong}, C., {Jin}, M., {Lingam}, M., {et~al.} 2018, Proceedings of the National
  Academy of Science, 115, 260, \dodoi{10.1073/pnas.1708010115}

\bibitem[{{Dotter}(2016)}]{dotter2016}
{Dotter}, A. 2016, \apjs, 222, 8, \dodoi{10.3847/0067-0049/222/1/8}

\bibitem[{{Driscoll} \& {Bercovici}(2013)}]{driscoll2013}
{Driscoll}, P., \& {Bercovici}, D. 2013, \icarus, 226, 1447,
  \dodoi{10.1016/j.icarus.2013.07.025}

\bibitem[{{Driscoll} \& {Barnes}(2015)}]{driscoll2015}
{Driscoll}, P.~E., \& {Barnes}, R. 2015, Astrobiology, 15, 739,
  \dodoi{10.1089/ast.2015.1325}

\bibitem[{{Duncan} {et~al.}(1998){Duncan}, {Levison}, \& {Lee}}]{duncan1998}
{Duncan}, M.~J., {Levison}, H.~F., \& {Lee}, M.~H. 1998, \aj, 116, 2067,
  \dodoi{10.1086/300541}

\bibitem[{{Fogg} \& {Nelson}(2009)}]{fogg2009}
{Fogg}, M.~J., \& {Nelson}, R.~P. 2009, \aap, 498, 575,
  \dodoi{10.1051/0004-6361/200811305}

\bibitem[{{Foley}(2015)}]{foley2015}
{Foley}, B.~J. 2015, \apj, 812, 36, \dodoi{10.1088/0004-637X/812/1/36}

\bibitem[{{Foley}(2019)}]{foley2019}
---. 2019, \apj, 875, 72, \dodoi{10.3847/1538-4357/ab0f31}

\bibitem[{{Foley} {et~al.}(2012){Foley}, {Bercovici}, \& {Landuyt}}]{foley2012}
{Foley}, B.~J., {Bercovici}, D., \& {Landuyt}, W. 2012, Earth and Planetary
  Science Letters, 331, 281, \dodoi{10.1016/j.epsl.2012.03.028}

\bibitem[{{Foley} \& {Smye}(2018)}]{foley2018a}
{Foley}, B.~J., \& {Smye}, A.~J. 2018, Astrobiology, 18, 873,
  \dodoi{10.1089/ast.2017.1695}

\bibitem[{{Gaidos} {et~al.}(2014){Gaidos}, {Mann}, {L{\'e}pine}, {Buccino},
  {James}, {Ansdell}, {Petrucci}, {Mauas}, \& {Hilton}}]{gaidos2014c}
{Gaidos}, E., {Mann}, A.~W., {L{\'e}pine}, S., {et~al.} 2014, \mnras, 443,
  2561, \dodoi{10.1093/mnras/stu1313}

\bibitem[{{Gaillard} \& {Scaillet}(2014)}]{gaillard2014}
{Gaillard}, F., \& {Scaillet}, B. 2014, Earth and Planetary Science Letters,
  403, 307, \dodoi{10.1016/j.epsl.2014.07.009}

\bibitem[{{Gallet} {et~al.}(2017){Gallet}, {Charbonnel}, {Amard}, {Brun},
  {Palacios}, \& {Mathis}}]{gallet2017a}
{Gallet}, F., {Charbonnel}, C., {Amard}, L., {et~al.} 2017, \aap, 597, A14,
  \dodoi{10.1051/0004-6361/201629034}

\bibitem[{{Georgakarakos} {et~al.}(2018){Georgakarakos}, {Eggl}, \&
  {Dobbs-Dixon}}]{georgakarakos2018}
{Georgakarakos}, N., {Eggl}, S., \& {Dobbs-Dixon}, I. 2018, \apj, 856, 155,
  \dodoi{10.3847/1538-4357/aaaf72}

\bibitem[{{Gilbert} {et~al.}(2021){Gilbert}, {Barclay}, {Quintana},
  {Walkowicz}, {Vega}, {Schlieder}, {Monsue}, {Cale}, {Collins}, {Gaidos}, {El
  Mufti}, {Reefe}, {Plavchan}, {Tanner}, {Wittenmyer}, {Wittrock}, {Jenkins},
  {Latham}, {Ricker}, {Rose}, {Seager}, {Vanderspek}, \& {Winn}}]{gilbert2021}
{Gilbert}, E.~A., {Barclay}, T., {Quintana}, E.~V., {et~al.} 2021, arXiv
  e-prints, arXiv:2109.03924.
\newblock \doarXiv{2109.03924}

\bibitem[{{Gladman}(1993)}]{gladman1993}
{Gladman}, B. 1993, \icarus, 106, 247, \dodoi{10.1006/icar.1993.1169}

\bibitem[{{Glaser} {et~al.}(2020){Glaser}, {Hartnett}, {Desch}, {Unterborn},
  {Anbar}, {Buessecker}, {Fisher}, {Glaser}, {Kane}, {Lisse}, {Millsaps},
  {Neuer}, {O'Rourke}, {Santos}, {Walker}, \& {Zolotov}}]{glaser2020a}
{Glaser}, D.~M., {Hartnett}, H.~E., {Desch}, S.~J., {et~al.} 2020, \apj, 893,
  163, \dodoi{10.3847/1538-4357/ab822d}

\bibitem[{{Hadden} \& {Lithwick}(2018)}]{hadden2018b}
{Hadden}, S., \& {Lithwick}, Y. 2018, \aj, 156, 95,
  \dodoi{10.3847/1538-3881/aad32c}

\bibitem[{{Hamano} {et~al.}(2013){Hamano}, {Abe}, \& {Genda}}]{hamano2013}
{Hamano}, K., {Abe}, Y., \& {Genda}, H. 2013, \nat, 497, 607,
  \dodoi{10.1038/nature12163}

\bibitem[{{Hayworth} \& {Foley}(2020)}]{hayworth2020b}
{Hayworth}, B. P.~C., \& {Foley}, B.~J. 2020, \apjl, 902, L10,
  \dodoi{10.3847/2041-8213/abb882}

\bibitem[{{Howe} {et~al.}(2020){Howe}, {Adams}, \& {Meyer}}]{howe2020}
{Howe}, A.~R., {Adams}, F.~C., \& {Meyer}, M.~R. 2020, \apj, 894, 130,
  \dodoi{10.3847/1538-4357/ab620c}

\bibitem[{{Johnstone} {et~al.}(2019){Johnstone}, {Khodachenko},
  {L{\"u}ftinger}, {Kislyakova}, {Lammer}, \& {G{\"u}del}}]{johnstone2019a}
{Johnstone}, C.~P., {Khodachenko}, M.~L., {L{\"u}ftinger}, T., {et~al.} 2019,
  \aap, 624, L10, \dodoi{10.1051/0004-6361/201935279}

\bibitem[{{Kaltenegger} {et~al.}(2010){Kaltenegger}, {Henning}, \&
  {Sasselov}}]{kaltenegger2010f}
{Kaltenegger}, L., {Henning}, W.~G., \& {Sasselov}, D.~D. 2010, \aj, 140, 1370,
  \dodoi{10.1088/0004-6256/140/5/1370}

\bibitem[{{Kane}(2014)}]{kane2014a}
{Kane}, S.~R. 2014, \apj, 782, 111, \dodoi{10.1088/0004-637X/782/2/111}

\bibitem[{{Kane}(2018)}]{kane2018a}
---. 2018, \apjl, 861, L21, \dodoi{10.3847/2041-8213/aad094}

\bibitem[{{Kane}(2019)}]{kane2019c}
---. 2019, \aj, 158, 72, \dodoi{10.3847/1538-3881/ab2a09}

\bibitem[{{Kane} \& {Deveny}(2018)}]{kane2018b}
{Kane}, S.~R., \& {Deveny}, S.~J. 2018, \apj, 864, 115,
  \dodoi{10.3847/1538-4357/aad802}

\bibitem[{{Kane} {et~al.}(2020{\natexlab{a}}){Kane}, {Fetherolf}, \&
  {Hill}}]{kane2020a}
{Kane}, S.~R., {Fetherolf}, T., \& {Hill}, M.~L. 2020{\natexlab{a}}, \aj, 159,
  176, \dodoi{10.3847/1538-3881/ab7818}

\bibitem[{{Kane} \& {Gelino}(2012)}]{kane2012a}
{Kane}, S.~R., \& {Gelino}, D.~M. 2012, \pasp, 124, 323, \dodoi{10.1086/665271}

\bibitem[{{Kane} {et~al.}(2021){Kane}, {Li}, {Wolf}, {Ostberg}, \&
  {Hill}}]{kane2021a}
{Kane}, S.~R., {Li}, Z., {Wolf}, E.~T., {Ostberg}, C., \& {Hill}, M.~L. 2021,
  \aj, 161, 31, \dodoi{10.3847/1538-3881/abcbfd}

\bibitem[{{Kane} {et~al.}(2018){Kane}, {Meshkat}, \& {Turnbull}}]{kane2018c}
{Kane}, S.~R., {Meshkat}, T., \& {Turnbull}, M.~C. 2018, \aj, 156, 267,
  \dodoi{10.3847/1538-3881/aae981}

\bibitem[{{Kane} {et~al.}(2020{\natexlab{b}}){Kane}, {Roettenbacher},
  {Unterborn}, {Foley}, \& {Hill}}]{kane2020d}
{Kane}, S.~R., {Roettenbacher}, R.~M., {Unterborn}, C.~T., {Foley}, B.~J., \&
  {Hill}, M.~L. 2020{\natexlab{b}}, The Planetary Science Journal, 1, 36,
  \dodoi{10.3847/PSJ/abaab5}

\bibitem[{{Kane} {et~al.}(2020{\natexlab{c}}){Kane}, {Turnbull}, {Fulton},
  {Rosenthal}, {Howard}, {Isaacson}, {Marcy}, \& {Weiss}}]{kane2020b}
{Kane}, S.~R., {Turnbull}, M.~C., {Fulton}, B.~J., {et~al.} 2020{\natexlab{c}},
  \aj, 160, 81, \dodoi{10.3847/1538-3881/ab9ffe}

\bibitem[{{Kane} {et~al.}(2020{\natexlab{d}}){Kane}, {Vervoort}, {Horner}, \&
  {Pozuelos}}]{kane2020e}
{Kane}, S.~R., {Vervoort}, P., {Horner}, J., \& {Pozuelos}, F.~J.
  2020{\natexlab{d}}, The Planetary Science Journal, 1, 42,
  \dodoi{10.3847/PSJ/abae63}

\bibitem[{{Kane} {et~al.}(2016){Kane}, {Hill}, {Kasting}, {Kopparapu},
  {Quintana}, {Barclay}, {Batalha}, {Borucki}, {Ciardi}, {Haghighipour},
  {Hinkel}, {Kaltenegger}, {Selsis}, \& {Torres}}]{kane2016c}
{Kane}, S.~R., {Hill}, M.~L., {Kasting}, J.~F., {et~al.} 2016, \apj, 830, 1,
  \dodoi{10.3847/0004-637X/830/1/1}

\bibitem[{{Kasting} \& {Catling}(2003)}]{kasting2003}
{Kasting}, J.~F., \& {Catling}, D. 2003, \araa, 41, 429,
  \dodoi{10.1146/annurev.astro.41.071601.170049}

\bibitem[{{Kasting} {et~al.}(1993){Kasting}, {Whitmire}, \&
  {Reynolds}}]{kasting1993a}
{Kasting}, J.~F., {Whitmire}, D.~P., \& {Reynolds}, R.~T. 1993, \icarus, 101,
  108, \dodoi{10.1006/icar.1993.1010}

\bibitem[{{Kite} \& {Barnett}(2020)}]{kite2020c}
{Kite}, E.~S., \& {Barnett}, M.~N. 2020, Proceedings of the National Academy of
  Science, 117, 18264.
\newblock \doarXiv{2006.02589}

\bibitem[{{Kleine} {et~al.}(2002){Kleine}, {M{\"u}nker}, {Mezger}, \&
  {Palme}}]{kleine2002}
{Kleine}, T., {M{\"u}nker}, C., {Mezger}, K., \& {Palme}, H. 2002, \nat, 418,
  952, \dodoi{10.1038/nature00982}

\bibitem[{{Kochukhov} \& {Reiners}(2020)}]{kochukhov2020b}
{Kochukhov}, O., \& {Reiners}, A. 2020, \apj, 902, 43,
  \dodoi{10.3847/1538-4357/abb2a2}

\bibitem[{{Kopparapu}(2013)}]{kopparapu2013b}
{Kopparapu}, R.~K. 2013, \apj, 767, L8, \dodoi{10.1088/2041-8205/767/1/L8}

\bibitem[{{Kopparapu} {et~al.}(2014){Kopparapu}, {Ramirez}, {SchottelKotte},
  {Kasting}, {Domagal-Goldman}, \& {Eymet}}]{kopparapu2014}
{Kopparapu}, R.~K., {Ramirez}, R.~M., {SchottelKotte}, J., {et~al.} 2014, \apj,
  787, L29, \dodoi{10.1088/2041-8205/787/2/L29}

\bibitem[{{Korenaga}(2010)}]{korenaga2010b}
{Korenaga}, J. 2010, \apjl, 725, L43, \dodoi{10.1088/2041-8205/725/1/L43}

\bibitem[{{Krissansen-Totton} \& {Catling}(2017)}]{krissansentotton2017}
{Krissansen-Totton}, J., \& {Catling}, D.~C. 2017, Nature Communications, 8,
  15423, \dodoi{10.1038/ncomms15423}

\bibitem[{{Krissansen-Totton} {et~al.}(2021){Krissansen-Totton}, {Galloway},
  {Wogan}, {Dhaliwal}, \& {Fortney}}]{krissansentotton2021b}
{Krissansen-Totton}, J., {Galloway}, M.~L., {Wogan}, N., {Dhaliwal}, J.~K., \&
  {Fortney}, J.~J. 2021, \apj, 913, 107, \dodoi{10.3847/1538-4357/abf560}

\bibitem[{{Lalitha} {et~al.}(2018){Lalitha}, {Schmitt}, \&
  {Dash}}]{lalitha2018}
{Lalitha}, S., {Schmitt}, J.~H.~M.~M., \& {Dash}, S. 2018, \mnras, 477, 808,
  \dodoi{10.1093/mnras/sty732}

\bibitem[{{Lammer} {et~al.}(2008){Lammer}, {Kasting}, {Chassefi{\`e}re},
  {Johnson}, {Kulikov}, \& {Tian}}]{lammer2008}
{Lammer}, H., {Kasting}, J.~F., {Chassefi{\`e}re}, E., {et~al.} 2008, \ssr,
  139, 399, \dodoi{10.1007/s11214-008-9413-5}

\bibitem[{{Lammer} {et~al.}(2009){Lammer}, {Bredeh{\"o}ft}, {Coustenis},
  {Khodachenko}, {Kaltenegger}, {Grasset}, {Prieur}, {Raulin}, {Ehrenfreund},
  {Yamauchi}, {Wahlund}, {Grie{\ss}meier}, {Stangl}, {Cockell}, {Kulikov},
  {Grenfell}, \& {Rauer}}]{lammer2009a}
{Lammer}, H., {Bredeh{\"o}ft}, J.~H., {Coustenis}, A., {et~al.} 2009, \aapr,
  17, 181, \dodoi{10.1007/s00159-009-0019-z}

\bibitem[{{Lenardic} \& {Crowley}(2012)}]{lenardic2012}
{Lenardic}, A., \& {Crowley}, J.~W. 2012, \apj, 755, 132,
  \dodoi{10.1088/0004-637X/755/2/132}

\bibitem[{{Leto} {et~al.}(2000){Leto}, {Pagano}, {Linsky}, {Rodon{\`o}}, \&
  {Umana}}]{leto2000}
{Leto}, G., {Pagano}, I., {Linsky}, J.~L., {Rodon{\`o}}, M., \& {Umana}, G.
  2000, \aap, 359, 1035

\bibitem[{{Levi} {et~al.}(2017){Levi}, {Sasselov}, \& {Podolak}}]{levi2017a}
{Levi}, A., {Sasselov}, D., \& {Podolak}, M. 2017, \apj, 838, 24,
  \dodoi{10.3847/1538-4357/aa5cfe}

\bibitem[{{Linsenmeier} {et~al.}(2015){Linsenmeier}, {Pascale}, \&
  {Lucarini}}]{linsenmeier2015}
{Linsenmeier}, M., {Pascale}, S., \& {Lucarini}, V. 2015, \planss, 105, 43,
  \dodoi{10.1016/j.pss.2014.11.003}

\bibitem[{{Luger} \& {Barnes}(2015)}]{luger2015b}
{Luger}, R., \& {Barnes}, R. 2015, Astrobiology, 15, 119,
  \dodoi{10.1089/ast.2014.1231}

\bibitem[{{MacGregor} {et~al.}(2013){MacGregor}, {Wilner}, {Rosenfeld},
  {Andrews}, {Matthews}, {Hughes}, {Booth}, {Chiang}, {Graham}, {Kalas},
  {Kennedy}, \& {Sibthorpe}}]{macgregor2013}
{MacGregor}, M.~A., {Wilner}, D.~J., {Rosenfeld}, K.~A., {et~al.} 2013, \apjl,
  762, L21, \dodoi{10.1088/2041-8205/762/2/L21}

\bibitem[{{Magee} {et~al.}(2003){Magee}, {G{\"u}del}, {Audard}, \&
  {Mewe}}]{magee2003}
{Magee}, H.~R.~M., {G{\"u}del}, M., {Audard}, M., \& {Mewe}, R. 2003, Advances
  in Space Research, 32, 1149, \dodoi{10.1016/S0273-1177(03)00321-1}

\bibitem[{{Mamajek} \& {Bell}(2014)}]{mamajek2014}
{Mamajek}, E.~E., \& {Bell}, C. P.~M. 2014, \mnras, 445, 2169,
  \dodoi{10.1093/mnras/stu1894}

\bibitem[{{Neves} {et~al.}(2013){Neves}, {Bonfils}, {Santos}, {Delfosse},
  {Forveille}, {Allard}, \& {Udry}}]{neves2013}
{Neves}, V., {Bonfils}, X., {Santos}, N.~C., {et~al.} 2013, \aap, 551, A36,
  \dodoi{10.1051/0004-6361/201220574}

\bibitem[{{Newton} {et~al.}(2019){Newton}, {Mann}, {Tofflemire}, {Pearce},
  {Rizzuto}, {Vanderburg}, {Martinez}, {Wang}, {Ruffio}, {Kraus}, {Johnson},
  {Thao}, {Wood}, {Rampalli}, {Nielsen}, {Collins}, {Dragomir}, {Hellier},
  {Anderson}, {Barclay}, {Brown}, {Feiden}, {Hart}, {Isopi}, {Kielkopf},
  {Mallia}, {Nelson}, {Rodriguez}, {Stockdale}, {Waite}, {Wright}, {Lissauer},
  {Ricker}, {Vanderspek}, {Latham}, {Seager}, {Winn}, {Jenkins}, {Bouma},
  {Burke}, {Davies}, {Fausnaugh}, {Li}, {Morris}, {Mukai}, {Villase{\~n}or},
  {Villeneuva}, {De Rosa}, {Macintosh}, {Mengel}, {Okumura}, \&
  {Wittenmyer}}]{newton2019}
{Newton}, E.~R., {Mann}, A.~W., {Tofflemire}, B.~M., {et~al.} 2019, \apjl, 880,
  L17, \dodoi{10.3847/2041-8213/ab2988}

\bibitem[{{Noack} \& {Breuer}(2014)}]{noack2014b}
{Noack}, L., \& {Breuer}, D. 2014, \planss, 98, 41,
  \dodoi{10.1016/j.pss.2013.06.020}

\bibitem[{{O'Brien} {et~al.}(2018){O'Brien}, {Izidoro}, {Jacobson}, {Raymond},
  \& {Rubie}}]{obrien2018}
{O'Brien}, D.~P., {Izidoro}, A., {Jacobson}, S.~A., {Raymond}, S.~N., \&
  {Rubie}, D.~C. 2018, \ssr, 214, 47, \dodoi{10.1007/s11214-018-0475-8}

\bibitem[{{O'Neill} \& {Lenardic}(2007)}]{oneill2007d}
{O'Neill}, C., \& {Lenardic}, A. 2007, \grl, 34, L19204,
  \dodoi{10.1029/2007GL030598}

\bibitem[{{Oosterloo} {et~al.}(2021){Oosterloo}, {H{\"o}ning}, {Kamp}, \& {van
  der Tak}}]{oosterloo2021}
{Oosterloo}, M., {H{\"o}ning}, D., {Kamp}, I.~E.~E., \& {van der Tak}, F.~F.~S.
  2021, \aap, 649, A15, \dodoi{10.1051/0004-6361/202039664}

\bibitem[{{Owen}(2019)}]{owen2019a}
{Owen}, J.~E. 2019, Annual Review of Earth and Planetary Sciences, 47, 67,
  \dodoi{10.1146/annurev-earth-053018-060246}

\bibitem[{{Paxton} {et~al.}(2011){Paxton}, {Bildsten}, {Dotter}, {Herwig},
  {Lesaffre}, \& {Timmes}}]{paxton2011}
{Paxton}, B., {Bildsten}, L., {Dotter}, A., {et~al.} 2011, \apjs, 192, 3,
  \dodoi{10.1088/0067-0049/192/1/3}

\bibitem[{{Paxton} {et~al.}(2013){Paxton}, {Cantiello}, {Arras}, {Bildsten},
  {Brown}, {Dotter}, {Mankovich}, {Montgomery}, {Stello}, {Timmes}, \&
  {Townsend}}]{paxton2013}
{Paxton}, B., {Cantiello}, M., {Arras}, P., {et~al.} 2013, \apjs, 208, 4,
  \dodoi{10.1088/0067-0049/208/1/4}

\bibitem[{{Paxton} {et~al.}(2015){Paxton}, {Marchant}, {Schwab}, {Bauer},
  {Bildsten}, {Cantiello}, {Dessart}, {Farmer}, {Hu}, {Langer}, {Townsend},
  {Townsley}, \& {Timmes}}]{paxton2015}
{Paxton}, B., {Marchant}, P., {Schwab}, J., {et~al.} 2015, \apjs, 220, 15,
  \dodoi{10.1088/0067-0049/220/1/15}

\bibitem[{{Paxton} {et~al.}(2018){Paxton}, {Schwab}, {Bauer}, {Bildsten},
  {Blinnikov}, {Duffell}, {Farmer}, {Goldberg}, {Marchant}, {Sorokina},
  {Thoul}, {Townsend}, \& {Timmes}}]{paxton2018}
{Paxton}, B., {Schwab}, J., {Bauer}, E.~B., {et~al.} 2018, \apjs, 234, 34,
  \dodoi{10.3847/1538-4365/aaa5a8}

\bibitem[{{Paxton} {et~al.}(2019){Paxton}, {Smolec}, {Schwab}, {Gautschy},
  {Bildsten}, {Cantiello}, {Dotter}, {Farmer}, {Goldberg}, {Jermyn}, {Kanbur},
  {Marchant}, {Thoul}, {Townsend}, {Wolf}, {Zhang}, \& {Timmes}}]{paxton2019}
{Paxton}, B., {Smolec}, R., {Schwab}, J., {et~al.} 2019, \apjs, 243, 10,
  \dodoi{10.3847/1538-4365/ab2241}

\bibitem[{{Pepe} {et~al.}(2000){Pepe}, {Mayor}, {Delabre}, {Kohler}, {Lacroix},
  {Queloz}, {Udry}, {Benz}, {Bertaux}, \& {Sivan}}]{pepe2000}
{Pepe}, F., {Mayor}, M., {Delabre}, B., {et~al.} 2000, in Society of
  Photo-Optical Instrumentation Engineers (SPIE) Conference Series, Vol. 4008,
  \procspie, ed. M.~{Iye} \& A.~F. {Moorwood}, 582--592,
  \dodoi{10.1117/12.395516}

\bibitem[{{Plavchan} {et~al.}(2020){Plavchan}, {Barclay}, {Gagn{\'e}}, {Gao},
  {Cale}, {Matzko}, {Dragomir}, {Quinn}, {Feliz}, {Stassun}, {Crossfield},
  {Berardo}, {Latham}, {Tieu}, {Anglada-Escud{\'e}}, {Ricker}, {Vanderspek},
  {Seager}, {Winn}, {Jenkins}, {Rinehart}, {Krishnamurthy}, {Dynes}, {Doty},
  {Adams}, {Afanasev}, {Beichman}, {Bottom}, {Bowler}, {Brinkworth}, {Brown},
  {Cancino}, {Ciardi}, {Clampin}, {Clark}, {Collins}, {Davison},
  {Foreman-Mackey}, {Furlan}, {Gaidos}, {Geneser}, {Giddens}, {Gilbert},
  {Hall}, {Hellier}, {Henry}, {Horner}, {Howard}, {Huang}, {Huber}, {Kane},
  {Kenworthy}, {Kielkopf}, {Kipping}, {Klenke}, {Kruse}, {Latouf}, {Lowrance},
  {Mennesson}, {Mengel}, {Mills}, {Morton}, {Narita}, {Newton}, {Nishimoto},
  {Okumura}, {Palle}, {Pepper}, {Quintana}, {Roberge}, {Roccatagliata},
  {Schlieder}, {Tanner}, {Teske}, {Tinney}, {Vanderburg}, {von Braun}, {Walp},
  {Wang}, {Wang}, {Weigand}, {White}, {Wittenmyer}, {Wright}, {Youngblood},
  {Zhang}, \& {Zilberman}}]{plavchan2020}
{Plavchan}, P., {Barclay}, T., {Gagn{\'e}}, J., {et~al.} 2020, \nat, 582, 497,
  \dodoi{10.1038/s41586-020-2400-z}

\bibitem[{{Ramirez} \& {Kaltenegger}(2014)}]{ramirez2014c}
{Ramirez}, R.~M., \& {Kaltenegger}, L. 2014, \apjl, 797, L25,
  \dodoi{10.1088/2041-8205/797/2/L25}

\bibitem[{{Raymond} {et~al.}(2006){Raymond}, {Mandell}, \&
  {Sigurdsson}}]{raymond2006d}
{Raymond}, S.~N., {Mandell}, A.~M., \& {Sigurdsson}, S. 2006, Science, 313,
  1413, \dodoi{10.1126/science.1130461}

\bibitem[{{Raymond} {et~al.}(2004){Raymond}, {Quinn}, \&
  {Lunine}}]{raymond2004a}
{Raymond}, S.~N., {Quinn}, T., \& {Lunine}, J.~I. 2004, \icarus, 168, 1,
  \dodoi{10.1016/j.icarus.2003.11.019}

\bibitem[{{Raymond} {et~al.}(2005){Raymond}, {Quinn}, \&
  {Lunine}}]{raymond2005b}
---. 2005, \icarus, 177, 256, \dodoi{10.1016/j.icarus.2005.03.008}

\bibitem[{{Raymond} {et~al.}(2012){Raymond}, {Armitage}, {Moro-Mart{\'\i}n},
  {Booth}, {Wyatt}, {Armstrong}, {Mand ell}, {Selsis}, \& {West}}]{raymond2012}
{Raymond}, S.~N., {Armitage}, P.~J., {Moro-Mart{\'\i}n}, A., {et~al.} 2012,
  \aap, 541, A11, \dodoi{10.1051/0004-6361/201117049}

\bibitem[{{Ribas} {et~al.}(2005){Ribas}, {Guinan}, {G{\"u}del}, \&
  {Audard}}]{ribas2005}
{Ribas}, I., {Guinan}, E.~F., {G{\"u}del}, M., \& {Audard}, M. 2005, \apj, 622,
  680, \dodoi{10.1086/427977}

\bibitem[{{Ricker} {et~al.}(2015){Ricker}, {Winn}, {Vanderspek}, {Latham},
  {Bakos}, {Bean}, {Berta-Thompson}, {Brown}, {Buchhave}, {Butler}, {Butler},
  {Chaplin}, {Charbonneau}, {Christensen-Dalsgaard}, {Clampin}, {Deming},
  {Doty}, {De Lee}, {Dressing}, {Dunham}, {Endl}, {Fressin}, {Ge}, {Henning},
  {Holman}, {Howard}, {Ida}, {Jenkins}, {Jernigan}, {Johnson}, {Kaltenegger},
  {Kawai}, {Kjeldsen}, {Laughlin}, {Levine}, {Lin}, {Lissauer}, {MacQueen},
  {Marcy}, {McCullough}, {Morton}, {Narita}, {Paegert}, {Palle}, {Pepe},
  {Pepper}, {Quirrenbach}, {Rinehart}, {Sasselov}, {Sato}, {Seager},
  {Sozzetti}, {Stassun}, {Sullivan}, {Szentgyorgyi}, {Torres}, {Udry}, \&
  {Villasenor}}]{ricker2015}
{Ricker}, G.~R., {Winn}, J.~N., {Vanderspek}, R., {et~al.} 2015, Journal of
  Astronomical Telescopes, Instruments, and Systems, 1, 014003,
  \dodoi{10.1117/1.JATIS.1.1.014003}

\bibitem[{{Rodr{\'\i}guez-Mozos} \& {Moya}(2019)}]{rodriguezmozos2019}
{Rodr{\'\i}guez-Mozos}, J.~M., \& {Moya}, A. 2019, \aap, 630, A52,
  \dodoi{10.1051/0004-6361/201935543}

\bibitem[{{Roettenbacher} \& {Kane}(2017)}]{roettenbacher2017}
{Roettenbacher}, R.~M., \& {Kane}, S.~R. 2017, \apj, 851, 77,
  \dodoi{10.3847/1538-4357/aa991e}

\bibitem[{{Schaefer} {et~al.}(2012){Schaefer}, {Lodders}, \&
  {Fegley}}]{schaefer2012}
{Schaefer}, L., {Lodders}, K., \& {Fegley}, B. 2012, \apj, 755, 41,
  \dodoi{10.1088/0004-637X/755/1/41}

\bibitem[{{Schlichting} {et~al.}(2015){Schlichting}, {Sari}, \&
  {Yalinewich}}]{schlichting2015}
{Schlichting}, H.~E., {Sari}, R., \& {Yalinewich}, A. 2015, \icarus, 247, 81,
  \dodoi{10.1016/j.icarus.2014.09.053}

\bibitem[{{Spurzem} {et~al.}(2009){Spurzem}, {Giersz}, {Heggie}, \&
  {Lin}}]{spurzem2009}
{Spurzem}, R., {Giersz}, M., {Heggie}, D.~C., \& {Lin}, D.~N.~C. 2009, \apj,
  697, 458, \dodoi{10.1088/0004-637X/697/1/458}

\bibitem[{{Strubbe} \& {Chiang}(2006)}]{strubbe2006}
{Strubbe}, L.~E., \& {Chiang}, E.~I. 2006, \apj, 648, 652,
  \dodoi{10.1086/505736}

\bibitem[{{Truitt} \& {Young}(2017)}]{truitt2017}
{Truitt}, A., \& {Young}, P.~A. 2017, \apj, 835, 87,
  \dodoi{10.3847/1538-4357/835/1/87}

\bibitem[{{Truitt} {et~al.}(2015){Truitt}, {Young}, {Spacek}, {Probst}, \&
  {Dietrich}}]{truitt2015}
{Truitt}, A., {Young}, P.~A., {Spacek}, A., {Probst}, L., \& {Dietrich}, J.
  2015, \apj, 804, 145, \dodoi{10.1088/0004-637X/804/2/145}

\bibitem[{{Tsikoudi} \& {Kellett}(2000)}]{tsikoudi2000b}
{Tsikoudi}, V., \& {Kellett}, B.~J. 2000, \mnras, 319, 1147,
  \dodoi{10.1046/j.1365-8711.2000.03905.x}

\bibitem[{{Underwood} {et~al.}(2003){Underwood}, {Jones}, \&
  {Sleep}}]{underwood2003}
{Underwood}, D.~R., {Jones}, B.~W., \& {Sleep}, P.~N. 2003, International
  Journal of Astrobiology, 2, 289, \dodoi{10.1017/S1473550404001715}

\bibitem[{{Valencia} {et~al.}(2007){Valencia}, {O'Connell}, \&
  {Sasselov}}]{valencia2007c}
{Valencia}, D., {O'Connell}, R.~J., \& {Sasselov}, D.~D. 2007, \apjl, 670, L45,
  \dodoi{10.1086/524012}

\bibitem[{{Valle} {et~al.}(2014){Valle}, {Dell'Omodarme}, {Prada Moroni}, \&
  {Degl'Innocenti}}]{valle2014b}
{Valle}, G., {Dell'Omodarme}, M., {Prada Moroni}, P.~G., \& {Degl'Innocenti},
  S. 2014, \aap, 567, A133, \dodoi{10.1051/0004-6361/201323350}

\bibitem[{{van Elteren} {et~al.}(2019){van Elteren}, {Portegies Zwart},
  {Pelupessy}, {Cai}, \& {McMillan}}]{vanelteren2019}
{van Elteren}, A., {Portegies Zwart}, S., {Pelupessy}, I., {Cai}, M.~X., \&
  {McMillan}, S.~L.~W. 2019, \aap, 624, A120,
  \dodoi{10.1051/0004-6361/201834641}

\bibitem[{{van Heck} \& {Tackley}(2011)}]{vanheck2011}
{van Heck}, H.~J., \& {Tackley}, P.~J. 2011, Earth and Planetary Science
  Letters, 310, 252, \dodoi{10.1016/j.epsl.2011.07.029}

\bibitem[{{Vogt} {et~al.}(1994){Vogt}, {Allen}, {Bigelow}, {Bresee}, {Brown},
  {Cantrall}, {Conrad}, {Couture}, {Delaney}, {Epps}, {Hilyard}, {Hilyard},
  {Horn}, {Jern}, {Kanto}, {Keane}, {Kibrick}, {Lewis}, {Osborne},
  {Pardeilhan}, {Pfister}, {Ricketts}, {Robinson}, {Stover}, {Tucker}, {Ward},
  \& {Wei}}]{vogt1994}
{Vogt}, S.~S., {Allen}, S.~L., {Bigelow}, B.~C., {et~al.} 1994, Society of
  Photo-Optical Instrumentation Engineers (SPIE) Conference Series, Vol. 2198,
  {HIRES: the high-resolution echelle spectrometer on the Keck 10-m Telescope}
  (SPIE Press), 362, \dodoi{10.1117/12.176725}

\bibitem[{{Walker} {et~al.}(1981){Walker}, {Hays}, \& {Kasting}}]{walker1981}
{Walker}, J.~C.~G., {Hays}, P.~B., \& {Kasting}, J.~F. 1981, \jgr, 86, 9776,
  \dodoi{10.1029/JC086iC10p09776}

\bibitem[{{Way} \& {Del Genio}(2020)}]{way2020}
{Way}, M.~J., \& {Del Genio}, A.~D. 2020, Journal of Geophysical Research
  (Planets), 125, e06276, \dodoi{10.1029/2019JE006276}

\bibitem[{{Way} \& {Georgakarakos}(2017)}]{way2017a}
{Way}, M.~J., \& {Georgakarakos}, N. 2017, \apj, 835, L1,
  \dodoi{10.3847/2041-8213/835/1/L1}

\bibitem[{{Wetherill}(1980)}]{wetherill1980a}
{Wetherill}, G.~W. 1980, \araa, 18, 77,
  \dodoi{10.1146/annurev.aa.18.090180.000453}

\bibitem[{{Wisdom}(2006)}]{wisdom2006b}
{Wisdom}, J. 2006, \aj, 131, 2294, \dodoi{10.1086/500829}

\bibitem[{{Wisdom} \& {Holman}(1991)}]{wisdom1991}
{Wisdom}, J., \& {Holman}, M. 1991, \aj, 102, 1528, \dodoi{10.1086/115978}

\bibitem[{{Young} {et~al.}(2012){Young}, {Liebst}, \& {Pagano}}]{young2012a}
{Young}, P.~A., {Liebst}, K., \& {Pagano}, M. 2012, \apjl, 755, L31,
  \dodoi{10.1088/2041-8205/755/2/L31}

\bibitem[{{Zendejas} {et~al.}(2010){Zendejas}, {Segura}, \&
  {Raga}}]{zendejas2010}
{Zendejas}, J., {Segura}, A., \& {Raga}, A.~C. 2010, \icarus, 210, 539,
  \dodoi{10.1016/j.icarus.2010.07.013}

\end{thebibliography}


\end{document}